\documentclass[aps,prd,reprint,superscriptaddress,nofootinbib,hyperref]{revtex4-1}
\pdfoutput=1
\usepackage{epsfig, subfigure}
\usepackage{setspace}
\usepackage{booktabs, tabularx} 
 
\usepackage{amsmath,bm,bbm}
\usepackage{color}

\usepackage{hyperref}
\usepackage{graphicx}
\usepackage{enumerate}
\hypersetup{
    pdfnewwindow=true,      
    colorlinks=true,       
    linkcolor=blue,          
    citecolor=blue,        
    filecolor=blue,      
    urlcolor=blue           
}

\def\lsim{\lesssim}  
\def\gsim{\gtrsim}

\newcommand{\beq}{\begin{equation}}  
\newcommand{\eeq}{\end{equation}}
\newcommand{\bea}{\begin{eqnarray}}  
\newcommand{\eea}{\end{eqnarray}}

\newcommand{\sigv}{\langle \sigma v\rangle}

\newcommand{\eg}{{\it e.g.}}
\newcommand{\etal }{{\it et al.}}

\begin{document}

\title{\mbox{\hspace{-0.6cm}Neutrinos in IceCube/KM3NeT as probes of Dark Matter Substructures in Galaxy Clusters}}

\author{Basudeb Dasgupta}
 \email{dasgupta.10@osu.edu}
 \affiliation{\mbox{Center for Cosmology and AstroParticle Physics, Ohio State University, 191 W.~Woodruff Ave., Columbus, 43210 OH, USA.}}

\author{Ranjan Laha}
\email{laha.1@osu.edu}
 \affiliation{\mbox{Center for Cosmology and AstroParticle Physics, Ohio State University, 191 W.~Woodruff Ave., Columbus, 43210 OH, USA.}}
\affiliation{Dept. of Physics, Ohio State University, 191 W.~Woodruff Ave., Columbus,  43210 OH, USA.}

 \date{\today}

\begin{abstract}
Galaxy clusters are one of the most promising candidate sites for dark matter annihilation. We focus on dark matter ($\chi$) with mass in the range $(10\,\rm{GeV}-100\,\rm{TeV})$, annihilating through the channels $\chi\chi\rightarrow\mu^+\mu^-$, $\chi\chi\rightarrow\nu\bar{\nu}$, $\chi\chi\rightarrow t \overline{t} $, or $\chi\chi\rightarrow\nu\bar{\nu}\nu \bar{\nu}$, and forecast the expected sensitivity to the annihilation cross section into these channels by observing galaxy clusters at IceCube/KM3NeT. Optimistically, the presence of dark matter substructures in galaxy clusters is predicted to enhance the signal by $(2-3)$ orders of magnitude over the contribution from the smooth component of the dark matter distribution. Optimizing for the angular size of the region of interest for galaxy clusters, the sensitivity to the annihilation cross section, $\langle \sigma v \rangle$, of heavy DM with mass in the range ($300\,{\rm GeV}-100\,{\rm TeV}$) will be $\mathcal{O}$($10^{-24}$\,cm$^3$s$^{-1}$), for full IceCube/KM3NeT live time of 10 years, which is about one order of magnitude better than the best  limit that can be obtained by observing the Milky Way halo. We find that neutrinos from cosmic ray interactions in the galaxy cluster, in addition to the atmospheric neutrinos, are a source of background. We show that significant improvement in the experimental sensitivity can be achieved for lower DM masses in the range ($10\,{\rm GeV}-300\,{\rm GeV})$ if neutrino-induced cascades can be reconstructed to $\approx 5^\circ$ accuracy, as may be possible in KM3NeT. We therefore propose that a low-energy extension ``KM3NeT-Core'', similar to DeepCore in IceCube, be considered for an extended reach at low DM masses.
\end{abstract}
\pacs{KM3NeT}

\maketitle

\section{Introduction}               \label{sec:Intro}

There is overwhelming evidence for, yet unexplained, invisible mass in our Universe~\cite{Zwicky:1933gu, Rubin:1970zz, Komatsu:2010fb, Clowe:2006eq}. Particles in the standard model of particle physics cannot account for the major fraction of this excess mass, but a new particle with weak-scale annihilation cross sections to standard model particles, as predicted in several extensions of the standard model of particle physics, would naturally explain its observed abundance~\cite{Jungman:1995df, Feng:2010gw, Bertone:2004pz, Bergstrom:2012np, Bertone:2010at}. This has motivated a comprehensive search for the particle identity of this ``dark matter'' (DM) using (i) direct production of DM at colliders~\cite{CMS, ATLAS}, (ii) direct detection of DM via elastic scattering~\cite{Bernabei:2008yi, Kim:2012rz, Aalseth:2010vx, Angloher:2011uu, Ahmed:2009zw, Ahmed:2010wy, Angle:2011th, Aprile:2011hi, Felizardo:2010mi, Behnke:2012ys, Armengaud:2011cy, Archambault:2012pm} and (iii) indirect detection of DM via its annihilation or decay~\cite{Gunn:1978gr,Stecker:1978du,Zeldovich:1980st,Ackermann:2011wa, IceCube:2011ae, Abdo:2010nc, Ajello:2011dq, Abdo:2010dk, Tanaka:2011uf, Abramowski:2010aa, Aleksic:2009ir, collaboration:2011sm, Aleksic:2011jx, Abbasi:2011eq, Adriani:2008zr, Adriani:2010rc, Ackermann:2010ij, FermiLAT:2011ab, Ackermann:2012qk, Aharonian:2008aa, Aharonian:2009ah}. This three-pronged approach to DM detection is necessary because a single experiment cannot probe all the properties of DM. For example, collider experiments mainly probe production of DM particles, whereas direct detection only probes the interaction between the DM particle and the particular detector material~\cite{Pato:2010zk}. Analogously, in an indirect detection experiment, we learn about the final states of DM annihilation or decay.

Indirect detection experiments are also sensitive to the DM density distribution at cosmological scales in this Universe unlike direct detection experiments which are only sensitive to the local DM distribution in the Milky Way. If LHC detects a DM candidate, then indirect detection experiments are also useful to determine whether that particular DM candidate makes up most of the DM in the Universe~\cite{Bertone:2011pq}. Indirect detection experiments, looking for products of DM annihilation in astrophysical sources, only detect a handful of the final states, \eg, photons, electrons, protons, neutrinos, and their antiparticles. If these experiments detect a signal that requires a cross section larger than the thermal relic annihilation cross section it would challenge a simple thermal WIMP paradigm of DM, and thus provide a crucial test of the WIMP paradigm~\cite{Steigman:2012nb}. On the other hand, there is no guarantee that a signal must be found if we can probe cross sections comparable to, or smaller than, the thermal relic annihilation cross section -- annihilations could proceed to undetected channels. In that case, however, one sets an upper bound on the partial annihilation cross sections into these observed channels, constraining particle physics models of DM.

Several astrophysical targets, \eg, the Sun, the Milky Way, dwarf galaxies, and galaxy clusters, may be observed by indirect detection experiments. A careful estimate of the signal and the background for each of these source classes is needed to determine which of these targets provides the best signal-to-noise ratio for a given DM model. The Sun accumulates DM particles while moving through the DM halo of the Milky Way. Due to the high density at the core of the Sun, for DM mass \,$\gsim$\,300 GeV, annihilations products are absorbed and the sensitivity of DM annihilation searches weakens considerably, making it inefficient for probing high mass DM~\cite{IceCube:2011aj, Rott:2011fh, Bell:2011sn, Taoso:2010tg, Lim:2009jy}. The Milky Way is dominated by DM in its central regions, but unknown astrophysical backgrounds make it difficult to disentangle the signal~\cite{Bertone:2004ag, Bertone:2002je, Weniger:2012tx, Bringmann:2012vr, Hooper:2011ti, Hooper:2010mq}, whereas the diffuse component of the Milky Way DM halo~\cite{Abdo:2010dk, Abbasi:2011eq} leads to a significantly reduced signal. Dwarf galaxies have a high mass-to-light ratio and are one of the ideal targets for detecting DM in gamma-ray experiments with subdegree angular resolution~\cite{IceCube:2011ae, PalomaresRuiz:2010pn, Ackermann:2011wa, GeringerSameth:2011iw, Belikov:2011pu, Buckley:2010vg, Aharonian:2007km, SanchezConde:2011ap, Sandick:2009bi, Moulin:2007ca}.

Galaxy clusters have the largest amount of DM amongst all known classes of gravitationally bound objects in the Universe. Although the background due to other astrophysical sources is also large therein, the contribution of DM substructures can enhance the DM annihilation signal from the smooth component, typically modeled using a Navarro, Frenk, and White (NFW) profile~\cite{Navarro:1996gj}. This enhancement depends on the abundance of DM substructures. State-of-the-art galaxy cluster simulations do not have the resolution to directly calculate the contribution due to the theoretically expected least massive substructures. However using theoretically well-motivated values for the mass of the smallest substructure and extrapolating the abundance of substructures to these lowest masses, high resolution computer simulations predict that galaxy clusters provide the best signal-to-noise ratio for DM annihilation signal~\cite{Gao:2011rf}. Note that even a moderate enhancement due to DM substructure, as advocated in~\cite{SanchezConde:2011ap} following the works in~\cite{Kamionkowski:2008vw,Kamionkowski:2010mi} predicts that galaxy clusters give the best signal-to-noise ratio for analysis where the field-of-view is greater than or equal to 1$^{\circ}$. This strongly motivates observations of galaxy clusters to search for DM annihilation signals\mbox{~\cite{Huang:2011xr, Abramowski:2012au, Dugger:2010ys, Profumo:2008fy, Ackermann:2010rg, Ando:2012vu, Pinzke:2009cp, Pinzke:2011ek, Jeltema:2008vu, Ackermann:2010rg}}.

Neutrino searches, among other indirect searches for DM, have distinct advantages. Being electrically neutral and weakly interacting, neutrinos travel undeflected and unattenuated from their sources. So neutrinos can provide information about dense sources, which may be at cosmological distances, from which no other standard model particles can reach us. Another crucial motivation to look for neutrinos is that many standard model particles eventually decay to produce neutrinos and gamma rays as final states. Detecting neutrinos is therefore complementary to gamma ray searches from DM annihilation, which have become very exciting in recent times~\cite{Ackermann:2011wa, Abdo:2010nc, Abramowski:2010aa, Abdo:2010dk, Aleksic:2011jx}. For very heavy DM, the gamma rays produced in the DM annihilations cascade and the constraints on DM annihilation cross sections become weaker than those obtained using neutrinos. Also, for hadronic explanations of any gamma ray and cosmic ray excesses, detecting neutrinos will be a smoking gun signature. Finally, direct annihilation to neutrinos is impossible to detect using any other detection channel, with electroweak bremsstrahlung being a notable exception~\cite{Bell:2011eu} although the limits obtained in that case turns out to be weaker than those obtained by direct observation of neutrinos~~\cite{Kachelriess:2007aj}. In fact, neutrinos, being least detectable, define a conservative upper bound on the DM annihilation cross section to standard model particles~\cite{Beacom:2006tt, Yuksel:2007ac}.

Limits obtained by gamma ray telescopes are typically stronger than that obtained using neutrino telescopes, but the larger angular resolution of a neutrino telescope, compared to a gamma ray telescope, means that the results obtained in a neutrino telescope is less dependent on the central part of the DM density profile (which gives the strongest signal in a gamma ray telescope) where the uncertainty obtained in DM simulations is the largest. Neutrinos telescopes are also able to view a target source for a longer time compared to a gamma ray telescope, though this advantage is mitigated by the smaller cross section of neutrino detection. Another advantage of neutrino telescopes is that they are able to view a large number of sources simultaneously and can be used to find dark matter in a region which is dark in the electromagnetic spectrum. These arguments and the availability of large neutrino telescopes strongly motivate a search for DM annihilation using neutrinos.

Although dwarf galaxies are known to be the best targets for dark matter searches for gamma-ray experiments, they are not the best targets for neutrino experiments. The reason for this is the limited angular resolution of a neutrino telescope, which is $\gsim$ 1$^{\circ}$. Dwarf galaxies have an angular size of $<$ 1$^{\circ}$ and thus when a neutrino telescope takes data from a dwarf galaxy, even with the minimum angular resolution, the size of the dwarf galaxy is smaller than the data-taking region, which implies a worse signal-to-noise ratio. However, galaxy clusters have a typical size of a few degrees and hence even when neutrino telescopes are taking data in the larger than minimum angular resolution mode, the size of the galaxy cluster fills up the entire data-taking region. This ensures that, unlike in the case of dwarf galaxies, there is no position in the data-taking region from where there is no potential signal candidate and thus provides a better signal-to-noise ratio.

Neutrinos from galaxy clusters have been considered previously by Yuan \etal\,\cite{Yuan:2010gn}.  In that paper, the DM halo for a galaxy cluster was obtained from extrapolation of the DM halo obtained from the simulation of a Milky Way like galaxy~\cite{Springel:2008by, Springel:2008cc}. Using the Fermi-LAT limits from galaxy clusters, Yuan \etal\,constrained the minimum DM substructure mass, and analyzed muon tracks in IceCube to obtain a constraint on DM annihilation cross section. 

In this paper, we investigate neutrinos from galaxy clusters using the latest DM density profiles, as given in Gao \etal\,\cite{Gao:2011rf}. This gives us updated inputs for both the smooth and the substructure components of DM in galaxy clusters. For comparison, we also calculate our results by taking the smooth and the substructure components of DM profile from the work by Sanchez-Conde \etal\,\cite{SanchezConde:2011ap} and find that due to the smaller boost factors (about a factor of 20 smaller boost factors than compared to that in~\cite{Gao:2011rf}), the sensitivity of the neutrino telescope for this parametrization of the DM profile is about a factor of 20 worse than what is obtained while using the DM substructure modeling of~\cite{Gao:2011rf}. We also take into account neutrinos produced due to cosmic ray interactions in the galaxy cluster, ignored in previous studies. With these updated inputs, we analyze the expected signals and backgrounds at IceCube and KM3NeT for both track and cascade events. While, quantitative improvement in the detection prospects is found for track searches, qualitative improvement in sensitivity and reach at low DM masses is expected if KM3NeT deploys a low energy extension, which we call KM3NeT-Core, and is able to reconstruct cascades with a pointing accuracy down to\,5$^\circ$ as claimed by Auer~\cite{Auer:2009zz}.

The remainder of the paper is arranged as follows. In Sec.\,\ref{sec:Cluster}, we discuss the neutrino flux from DM annihilation, using the DM density profile of a typical galaxy cluster. In Sec.\,\ref{sec:NeuDet}, we discuss neutrino detection and relevant backgrounds at a neutrino telescope. In Sec.\,\ref{sec:Results} we discuss the results, showing our forecasted sensitivity to $\sigv$ for the considered annihilation channels, and conclude in Sec.\,\ref{sec:Conclusion}.

\section{DM Distribution and Neutrino Production in Galaxy Clusters}            \label{sec:Cluster}

The  number flux of neutrinos per unit energy interval (in $\rm GeV^{-1} cm^{-2} s^{-1}$) for a given final state of DM annihilation is given by~\cite{Ando:2012vu}
\begin{equation}
\label{eq:number flux of neutrinos}
\frac{d\Phi_{\nu}}{dE_{\nu}}=\int _{\Delta \Omega} d\Omega\dfrac{1}{8 \pi m_{\chi}^2} \langle \sigma v \rangle \frac{dN_{\nu}}{dE_{\nu}} \int dl \mbox{ }\rho^2[r(l),\psi]~,
\end{equation}
where  $m_{\chi}$ denotes the mass of the DM particle (in units of GeV), $\langle \sigma v \rangle$ denotes the thermal-averaged annihilation cross section into the final state which can produce neutrinos (in units of $\rm cm^3s^{-1}$). $dN_{\nu}/dE_{\nu }$ denotes the energy spectrum  of the neutrinos from the various final states of DM annihilation (in units of ${\rm GeV^{-1}}$). The integral $\int dl\,\rho^2[r(l),\psi]$ is the line-of-sight integral of the DM density distribution, with $l$ denoting the line-of-sight distance (in units of cm), $\rho(r)$ denoting the DM density distribution function at a point $r$ (in units of GeVcm$^{-3}$). 

We have assumed here that the DM is its own anti-particle, which gives an extra factor of 2 in the denominator of the expression in Eq.\,(\ref{eq:number flux of neutrinos}). We also assume that the galaxy cluster is close enough so that the neutrino energy is not red-shifted significantly. We emphasize that even for a neutrino telescope, a nearby galaxy cluster is not a point source and hence an angular dependence, $\psi$, of the line-of-sight integral is present. Therefore, we have to integrate over the relevant solid angle, $\Delta \Omega = 2\pi \int_0 ^{\psi_{\rm{max}}} \sin\psi\,d\psi$, where $\psi_{\rm{max}}$ is the angular radius of the region of interest.

It can be seen that the neutrino flux is written in such a way that it is a product of the astrophysics quantities, $\int dl\,\rho^2[r(l),\psi]$, with the particle physics quantities, $m_{\chi}^{-2}(\sigv/2)dN_{\nu}/dE_{\nu}$. In the following subsections, we outline how we have calculated each of these quantities for our analysis.

\subsection{DM distribution}
In this section, we describe the DM density distribution in a typical galaxy cluster. Although we shall refer to the Virgo galaxy cluster for specific quantitative details, the same physical description is qualitatively applicable to other galaxy clusters.

\begin{figure}[!t]
\includegraphics[width=1 \columnwidth]{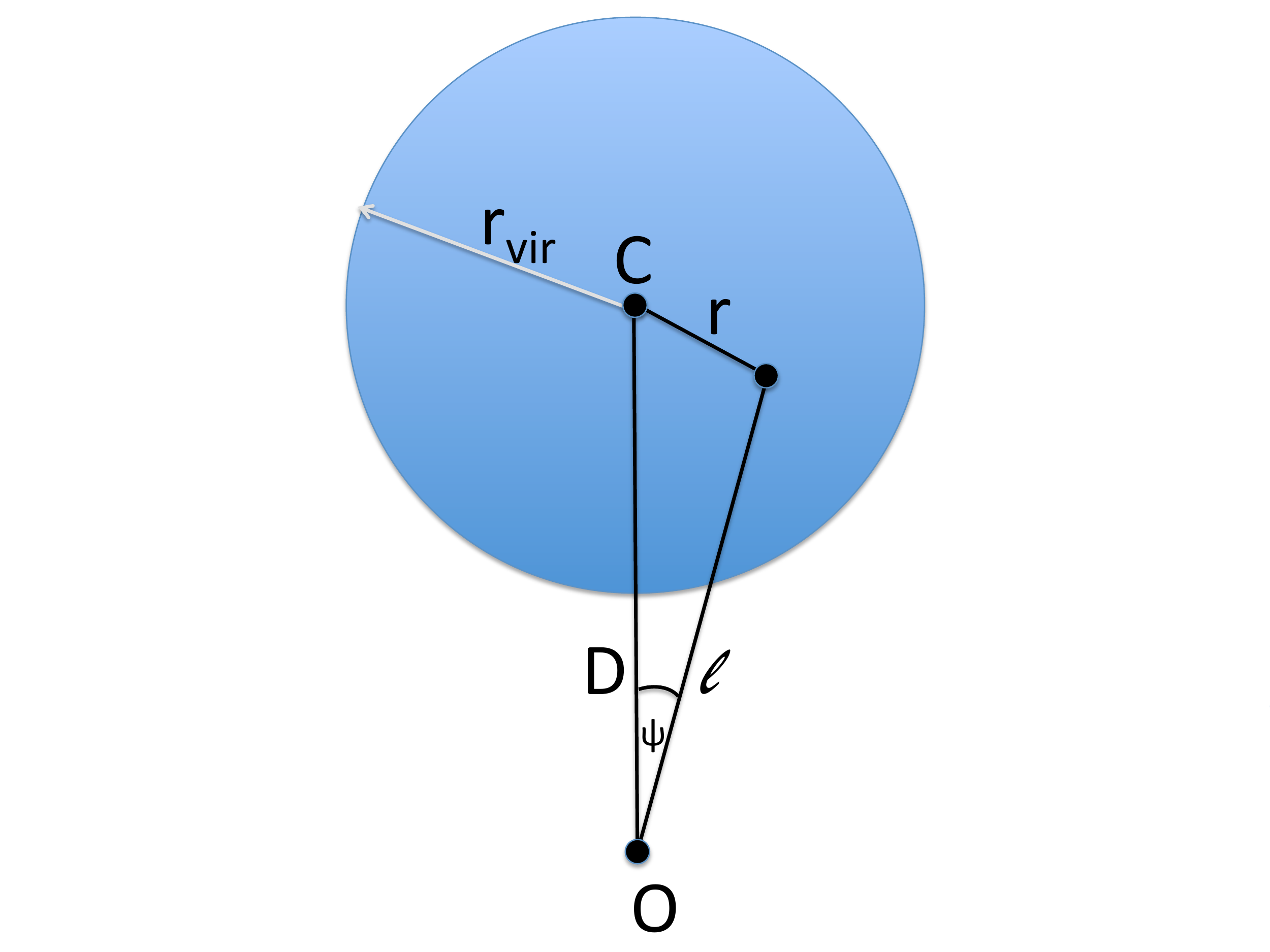}
\caption{Figure to illustrate the line-of-sight integral. The blue shaded region is the galaxy cluster with $C$ as its centre. The position of the observer is marked by the point $O$. The virial radius of the galaxy cluster is denoted by $r_{\rm {vir}}$. The distance of the observer, $O$, to the centre of the galaxy cluster, $C$, is denoted by $D$. The distance of a typical point inside the galaxy cluster from the centre of the galaxy cluster and the observer is denoted by $r$ and $l$ respectively.}
\label{fig:LOS integral}
\end{figure}

Galaxy clusters are the most massive gravitationally bound objects in the Universe today~\cite{Kravtsov:2012zs, Voit:2004ah}. A typical galaxy cluster has a mass of $\sim$~$\mathcal{O}(10^{14}\,M_{\odot}$) and is virialized up to a radius of $\sim$~$\mathcal{O}$(Mpc). We take the smooth component of the DM density profile to be parametrized by an NFW profile~\cite{Navarro:1996gj}
\begin{equation}
\label{eq:NFW}
\rho (r)=\dfrac{\rho_s}{\left(\dfrac{r}{r_s}\right)\left(1+\dfrac{r}{r_s}\right)^2} ~,
\end{equation}
 where $r_s$ is the scale radius and $\rho_s$ is the DM density at distance $\sim\mathcal{O}(r_s)$ from the centre of the galaxy cluster. 
 
 Given the redshift, $z$, and  the virial mass, $M_{\rm{vir}}$, of a galaxy cluster, the virial radius, $r_{\rm{vir}}$, can be determined from the following relationship, as given by Ando and Nagai~\cite{Ando:2012vu},
\begin{equation}
\label{eq:determine rvir}
M_{\rm{vir}}=\dfrac{4}{3}\pi r_{\rm{vir}}^3\Delta_{\rm{vir}}(z)\rho_c(z)~.
\end{equation}
Here, virial quantities are identified by using ``vir" in the subscript, $\rho_c(z)$ is the critical density of the Universe and the cosmological factor $\Delta_{\rm{vir}}(z)$ $=$ $82d -39d^2+18\pi^2$, where $d=-\Omega_{\Lambda}/\left(\Omega_{\Lambda}+\Omega_m\left(1+z\right)^3\right)$~\cite{Bryan:1997dn}. We assume a $\Lambda$CDM model for the Universe for all calculations: $\Omega_{\Lambda}$ = 0.73, $\Omega_{\rm m}$ = 0.27 and $H_0$ = 73 km s$^{-1}$ Mpc$^{-1}$~\cite{Nakamura:2010zzi}. The scale radius is obtained from the equation $r_s=r_{\rm vir}/c_{\rm vir}$, where $c_{\rm vir}$ denotes the concentration parameter which is given by~\cite{Ando:2012vu},
\begin{equation}
\label{eq:determine cvir}
c_{\rm{vir}}=\dfrac{7.85}{(1+z)^{0.71}}\left(\dfrac{M_{\rm{vir}}}{2\times 10^{12}h^{-1}M_{\odot}} \right)^{-0.081}~.
\end{equation}
 To obtain $\rho_s$, we equate the virial mass of the cluster, $M_{\rm vir}$, to the volume integral of $\rho(r)$ up to $r_{\rm vir}$. Analytically, we obtain
\begin{equation}
\label{eq:analytical rhos}
\rho_s=\dfrac{\Delta_{\rm{vir}}}{3}\dfrac{c_{\rm{vir}}^3\rho_c}{{\rm log}(1+c_{\rm{vir}})-\dfrac{c_{\rm{vir}}}{1+c_{\rm{vir}}}}
\end{equation}
 
For the Virgo galaxy cluster, the virial mass is taken to be $M_{\rm vir}=6.9\,\times$\,$10^{14}$\,$M_{\odot}$~\cite{Fouque:2001qc} and the redshift is taken to be $z=0.0036$~\cite{ned}. Note that the redshift is too small to affect neutrino energies appreciably. Using the value of the critical density of the Universe, $\rho_c=\rm 0.54 \times 10^{-5}\,GeV\,cm^{-3}$~\cite{Nakamura:2010zzi}, we get the virial radius of the Virgo galaxy cluster to be $r_{\rm vir}$\,=\,2.29\,Mpc. We use the concentration parameter, $c_{\rm{vir}}=4.98$, as in~\cite{Ando:2012vu}, which gives $r_s=0.46$\,Mpc. For the Virgo galaxy cluster, we find that $\rho_s$ = 2.19\,$\times$\,10$^{-2}$\,$\rm{GeV}\,\rm{cm}^{-3}$. We note that this value of the central DM density is about a factor of 2 lower than what we would have obtained if we had followed the prescription in Han \etal ~\cite{Han:2012au}. This difference can be traced to the fact that the cosmological factor, $\Delta_{\rm{vir}}(z)$, in our calculation has a value of $\sim$ 100, whereas the similar expression for $\rho_s$, as given in~\cite{Han:2012au}, gives the cosmological factor to be 200. We adopt the optimistic value of $\Delta_{\rm{vir}}$, and hence use $\rho_s$ = 4.38\,$\times$\,10$^{-2}$\,$\rm{GeV}\,\rm{cm}^{-3}$ throughout this work, but remind the readers that a lower value of $\Delta_{\rm {vir}}(z)$ by a factor of two can decrease the annihilation signal and sensitivity by a factor of 4.

In Fig.\,\ref{fig:LOS integral}, we schematically show how to calculate the line-of-sight integral.
Here $O$ is the position of the observer and the blue shaded region is the galaxy cluster whose centre is denoted by $C$. The virial radius of the galaxy cluster is shown as $r_{\rm{vir}}$ and the distance to the centre of the galaxy cluster is denoted by $D$. The line-of-sight distance to a point inside the galaxy cluster which is at a distance $r$ from the centre of the galaxy cluster is given by $l$. 

The line-of-sight integral, as a function of the angle $\psi$, is defined as  
\begin{eqnarray}
\label{eq:expression for J_NFW}
j(\psi)=\int _{l_{\rm {min}}}^{l_{\rm{max}}}dl \mbox{ }\rho^2[r(l),\psi]\,,
\end{eqnarray}
where
\bea
r&=&\sqrt{l^2+D^2-2Dl\cos \psi}\,,\\
l_{\rm{max,min}}&=&D\cos \psi ~\pm ~\sqrt{D^2\cos^2\psi-\left(D^2-r^2_{\rm{vir}}\right)}\,.
\label{eq:l_min and l_max}
\end{eqnarray}
We call this integral the $j$-factor for future reference. The distance to Virgo galaxy cluster is taken to be $D=19.4$\,Mpc~\cite{ned}. Using the parameters mentioned above, we find that
\begin{eqnarray}
\label{eq:Jsolidangle}
J_{\rm NFW}(\psi_{\rm max})&=&\int _0 ^{\Delta \Omega} d\Omega \, j_{\rm{NFW}}(\psi) \nonumber\\ 
 &=& 2.064 \times 10^{-6} ~\mbox{GeV}^2\rm{cm}^{-6}\rm{Mpc}\,,
\end{eqnarray}
where $\Delta \Omega = 2\pi \int_0 ^{\psi_{\rm{max}}} \sin\psi\,d\psi$, and $\psi_{\rm max}\approx\,6^{\circ}$ for the Virgo galaxy cluster.  
Recent high resolution simulations of galaxy clusters, in particular the Phoenix project~\cite{Gao:2011rf}, show a high concentration of DM substructures in addition to the smooth NFW profile. Tidal forces destroy the smallest mass substructures in the inner regions of the galaxy cluster so the inner region of a galaxy cluster ($\lsim$\,1 kpc) is dominated by the smooth NFW profile. However, the DM density in the outer region of a galaxy cluster is dominated by the DM substructures~\cite{Gao:2011rf}. This suggests that one should search for extended emission while looking for DM annihilation signal from a galaxy cluster.

The contribution to the DM annihilation due to substructures depends on their abundance. Recent simulations can only resolve substructures  of masses $\gtrsim 10^{5}\,M_{\odot}$ but theoretical considerations suggest that the minimum substructure mass for cold DM is in the range $(10^{-12}\,M_{\odot}-10^{-6}\,M_{\odot})$~\cite{Bringmann:2009vf}. In order to obtain the DM annihilation signal, we have to extrapolate the substructure abundance, using a halo mass distribution function from the simulations, from a mass of $\sim 10^5\,M_{\odot}$ to a minimum substructure mass of $\sim 10^{-6}\,M_{\odot}$. This 11 orders of magnitude extrapolation is the largest source of uncertainty in our calculation. However, it must be emphasized that even with a mass resolution of~\mbox{$\sim5\,\times 10^{7}\,M_{\odot}$}, the galaxy cluster simulations predict that the substructure contribution completely dominates the smooth contribution at radii $\gsim\,400$\,kpc~\cite{Gao:2011rf}. 

Assuming the smallest substructures to have masses $\sim10^{-6}\,M_{\odot}$, Han \etal\,\cite{Han:2012au} parametrize the $j$-factor due to substructures as
\begin{eqnarray}
\label{eq:Jsubstructure for angle less than psi_200}
j_{\rm{sub}}(\psi)\bigg|_{\psi \leq \psi_{200}}=\frac{b(M_{200}) J_{\rm NFW}}{\pi ~\mbox{ln}\,17}\frac{1}{\psi^2~+~(\psi/4)^2}
\end{eqnarray}
and 
\begin{eqnarray}
\label{eq:Jsubstructure for angle greater than psi_200}
j_{\rm{sub}}(\psi)\bigg|_{\psi \geq \psi_{200}}=j_{\rm{sub}}(\psi_{200})~e^{-2.377\left(\frac{\psi-\psi_{200}}{\psi_{200}}\right)},
\end{eqnarray}
where $b(M_{200})=1.6\times 10^{-3}(M_{200}/M_{\odot})^{0.39}$ is the boost factor. Here $M_{200}$, $\psi_{200}$, and $r_{200}$ are the mass, angular radius, and radius of the cluster where the average DM density is 200 times the critical density of the Universe. Using the value of $M_{200}$, as given in~\cite{Han:2012au}, we obtain the boost factor, $b(M_{200})\,\approx$\,980. As mentioned in~\cite{Gao:2011rf}, this boost factor is about an order of magnitude larger than the analogous boost factor obtained from galaxy halos. A boost factor of $\sim$ 1000 for galaxy clusters was also analytically obtained in~\cite{Pinzke:2011ek}.

Here we again mention that if we follow the galaxy cluster DM substructure modeling of~\cite{SanchezConde:2011ap}, the boost factor that we obtain is 55 for the Virgo galaxy cluster and between 34 and 54 for other galaxy clusters that were considered in that work. Hence there is a factor of $\sim$20 uncertainty in the sensitivity to DM particle properties that can be derived from observation of galaxy clusters both by gamma-rays observations and neutrino observations.

\begin{figure}[!ts]
\includegraphics[width=1 \columnwidth]{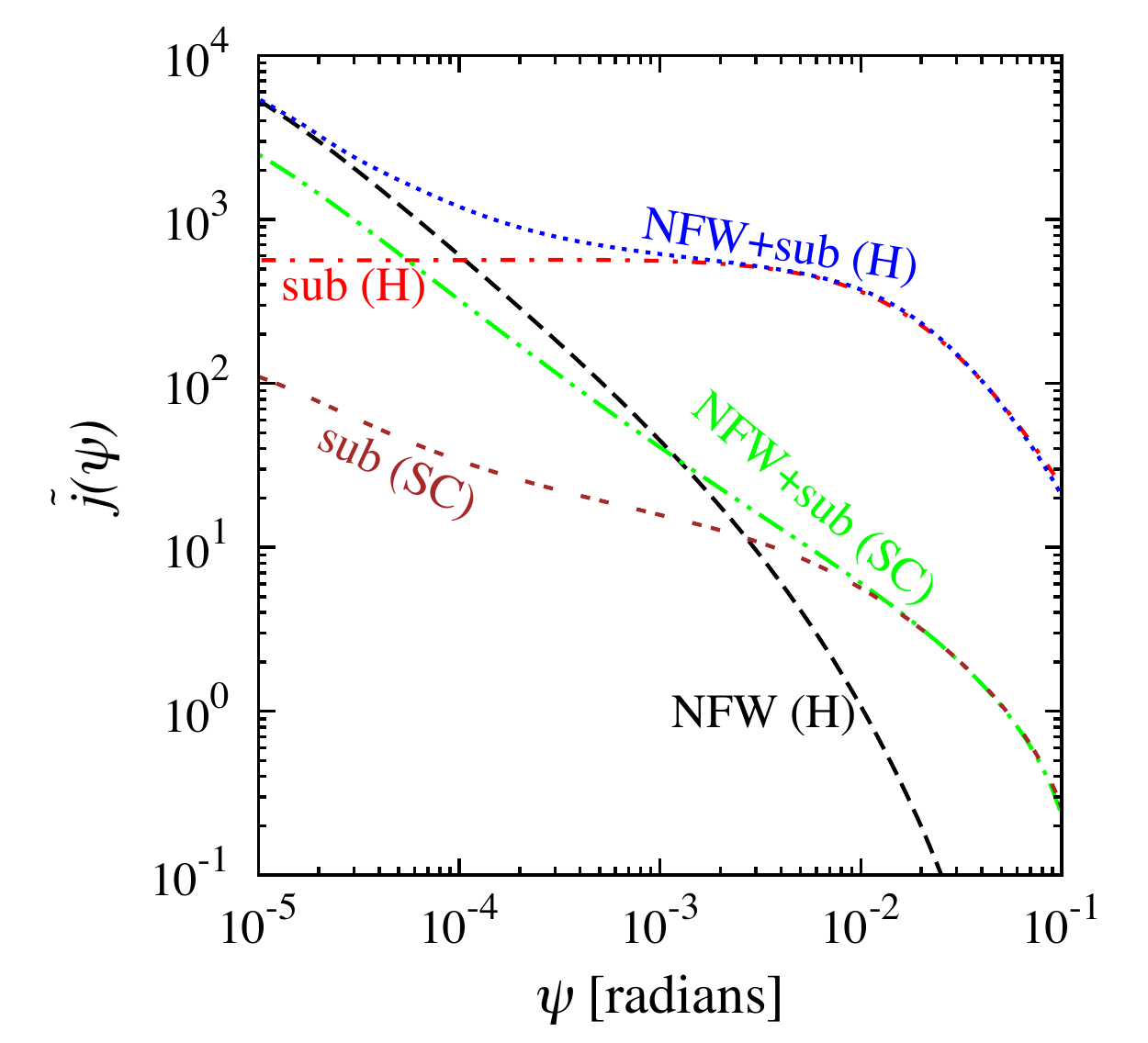}
\caption{Scaled line-of-sight integral (l.o.s.), $\tilde{j}(\psi)$, as a function of the angle $\psi$ (see Eq.\,\ref{eq:scaled J}). The black line shows the scaled l.o.s. for an NFW halo. The red dot-dashed line represents the scaled l.o.s. due to DM substructures following the work of~\cite{Han:2012au}. We also show the scaled l.o.s. due to DM substructures following the work of~\cite{SanchezConde:2011ap} by the brown dash-dash-gap line. The blue dotted line is the combined contribution of the NFW halo and the substructure following the work of~\cite{Han:2012au}. The combined contribution of the NFW halo and the substructure following the work of~\cite{SanchezConde:2011ap} is also shown by the green dash-dot-dot line. H in parenthesis denotes parametrization taken from~\cite{Han:2012au} where SC in parenthesis denotes parametrization taken from~\cite{SanchezConde:2011ap}. Figure adapted from~\cite{Han:2012au}.}
\label{fig:J vs psi}
\end{figure}

We scale the line-of-sight integral $j(\psi)$ to our local DM density-squared and to the distance to the Galactic centre from the Sun to define the scaled $j$-factor
\begin{eqnarray}
\label{eq:scaled J}
\tilde{j}(\psi)=\int \frac{dl}{8.5\,{\rm kpc}} \left(\frac{\rho[r(l),\psi]}{0.3\,{\rm GeVcm^{-3}}}\right)^2\,.
\end{eqnarray}
In Fig.\,\ref{fig:J vs psi}, we plot $\tilde{j}(\psi)$ against angle $\psi$, for the Virgo galaxy cluster for both the DM profile models in~\cite{Han:2012au,SanchezConde:2011ap}. It is easily seen that the presence of substructure provides a large boost to the DM annihilation signal for both the DM profile models, although the boost factors are different for both the models. The contribution from the NFW halo is concentrated at the centre whereas the contribution from the DM substructure is more extended for both the DM profile models. We use the model in~\cite{Han:2012au} for all our subsequent results. To obtain the results for the DM profile modeling with~\cite{SanchezConde:2011ap}, one can simply decrease the sensitivity in the result section~\ref{sec:Results} by a factor of $\sim$20. We remind the reader that at present due to the limited numerical resolution of the DM simulations, it is impossible to completely resolve the question of the boost factor which not only depends on the lowest DM substructure mass but also on merging of different galaxies to form a galaxy cluster. 

\subsection{Neutrino spectrum at source}
Now we turn our attention to the particle physics relevant for calculating the neutrino flux from DM annihilation. Since the DM in galaxy clusters is non-relativistic, with typical velocities $v\sim~10^{3}\,{\rm km\,s^{-1}}$, the DM annihilation products in a 2-body final state with identical particles are produced with an energy equal to the mass of the DM particle.

In this paper we study the the sensitivity of neutrino telescopes to $\sigv$ for DM annihilation to four interesting channels: (i) $\chi \chi \rightarrow \mu^+ \mu^-$, (ii) $\chi \chi \rightarrow \nu \overline{\nu}$, (iii) $\chi \chi \rightarrow~t \overline{t}$ and (iv)\,$\chi \chi \rightarrow VV \rightarrow\nu \overline{\nu} \nu \overline{\nu}$. All these chosen final state particles have or produce neutrinos on decay, and we forecast the sensitivity to the annihilation cross section that can be obtained using a neutrino telescope.

\subsubsection{$\chi \chi \, \rightarrow \, \mu^+\mu^-$}
The $\chi \chi \rightarrow \mu^+ \mu^-$ channel leads to signals in both gamma rays and neutrinos, and therefore quite promising for multi messenger studies. 
The normalised neutrino spectrum from decays of energetic muons of energy $E_{\mu}$ is given by~\cite{Lipari:1993hd}:
\begin{eqnarray}
\label{eq:numu neutrino spectrum from energetic muon decay}
\frac{dN_{\nu_{\mu}}}{dE_{\nu_{\mu}}}=\frac{5}{3E_{\mu}}-\frac{3E_{\nu_{\mu}}^2}{E_{\mu}^3}+\frac{4E_{\nu_{\mu}}^3}{3E_{\mu}^4} ~,
\end{eqnarray}
and 
\begin{eqnarray}
\label{eq:nue neutrino spectrum from energetic muon decay}
\frac{dN_{\overline{\nu}_e}}{dE_{\overline{\nu}_e}}=\frac{2}{E_{\mu}}-\frac{6E_{\overline{\nu}_e}^2}{E_{\mu}^3}+\frac{4E_{\overline{\nu}_e}^3}{E_{\mu}^4} ~.
\end{eqnarray}
Neutrino oscillations ensure that there is a 1:1:1 ratio of the fluxes of the $\nu_e$, $\nu_{\mu}$ and $\nu_{\tau}$ reaching the detector. 
An analogous equation holds true for antineutrinos.

\subsubsection{$\chi \chi \, \rightarrow \, \nu\overline{\nu}$}
Searching for direct annihilation to neutrinos is motivated by the presence of sharp spectral feature in the neutrino spectrum in the channel. Although this channel is suppressed for a Majorana or a scalar DM particle, there exist models in which the DM coupling to neutrinos is enhanced. This channel also gives the most stringent limits for DM annihilation in a neutrino telescope.

The neutrino spectrum due to direct annihilation to neutrinos is given by 
\begin{eqnarray}
\label{eq:neutrino spectrum from DM annihilation to neutrinos}
\frac{dN_{\nu}}{dE_{\nu}}=\delta(E_{\nu}-m_{\chi})~. 
\end{eqnarray}
Due to the finite energy resolution of the neutrino telescope, the dirac-delta function gets smeared out. We model the neutrino spectrum as a gaussian with centre at $m_{\chi}$ and a full-width at half-maximum given by the energy resolution of the detector~\cite{Resconi:2008fe}. Neutrino oscillations ensure that there is a 1:1:1 ratio of the fluxes of the $\nu_e$, $\nu_{\mu}$ and $\nu_{\tau}$ reaching the detector. 

\subsubsection{$\chi \chi \, \rightarrow \, t\overline{t}$}
The third channel which we consider is $\chi \chi \rightarrow t \overline{t}$. This is the most favored annihilation channel, from helicity arguments, for heavy ($\gsim$\,175~GeV) DM, if the DM is a Majorana fermion or a scalar.

The top quark decays to $W$-boson and a $b$-quark with a branching ratio of $\gsim$\,99\% and the subsequent decay of $W$-boson and hadronization of the $b$-quark produces neutrinos.  As an approximation, we consider only the prompt neutrinos produced by the decay of the $W$-boson and the $b$-quark. Following ~\cite{Jungman:1994jr}, we derive the highest energy muon neutrino flux due to the top quark decay as 
\begin{eqnarray}
\label{eq:neutrino spectrum due to top decay}
\frac{dN_{\nu}}{dE_{\nu}}&=&\frac{1}{3}\bigg(\sum _l \frac{\Gamma _{W \rightarrow l \nu_l}}{2 \gamma_t \beta_t E_W \beta_W} \rm{ln}\frac{\rm{max}(E_+, \epsilon_+)}{\rm{min}(E_-, \epsilon_-)}  \bigg) \nonumber\\
&&\times\Theta(E_{\nu}-\gamma_t(1-\beta_t)\epsilon_-) \times \Theta(\gamma_t(1+\beta_t)\epsilon_+-E_{\nu}) \nonumber\\
&+&\frac{1}{3} \bigg(\sum _l \frac{\Gamma _{b \rightarrow l \nu_l X}}{2\gamma_t E_d \beta_t}D_b\bigg[\frac{E_-}{E_d},\mbox{min}\bigg(1,\frac{E_+}{E_d}\bigg)\bigg]\bigg) \nonumber\\
&&\times\Theta(\gamma_t(1+\beta_t)E_d-E_{\nu})\,,
\end{eqnarray}
where $l$ denotes the relevant lepton states in the decay of the $W$-boson and the decay of the $b$-hadrons. The corresponding branching ratio for the decay of the $W$-boson and the $b$-hadrons is denoted by $\Gamma$, and the corresponding values are  taken from PDG~\cite{Nakamura:2010zzi}. The Lorentz factor is denoted by $\gamma_t=E_t/m_t=1/\sqrt{1-\beta_t^2}$. E$_W$ and $\beta_W$ are the energy and velocity of the $W$-boson in the top quark rest frame. $E_\pm=E_{\nu}\gamma_t^{-1}/(1\mp \beta_t)$ represents the maximum and minimum energy of the neutrino in the moving frame of the top quark. The limits of the neutrino energy in the moving frame of the $W$-boson is denoted by $\epsilon_\pm=E_W(1\pm \beta_W)/2$. If the energy of the $b$-quark in the rest frame of the top quark is denoted by $E_b$ then the hadronization energy can be approximated as $E_d=z_f E_b$ where we take the value of $z_f$ from~\cite{Jungman:1994jr}. The function $D_b[x,y]=\frac{1}{3}\left(6 \mbox{ ln}(y/x)+4(y^3-x^3)+9(x^2-y^2)\right)$. We ignore the lower energy muon neutrinos produced due to the decay of the muons in the final state. Neutrino oscillations ensure that there is a 1:1:1 ratio of the fluxes of the $\nu_e$, $\nu_{\mu}$ and $\nu_{\tau}$ reaching the detector.

\subsubsection{$\chi \chi \, \rightarrow \,VV \rightarrow \nu\overline{\nu} \nu \overline{\nu}$}
This channel is motivated by the secluded DM models~\mbox{\cite{Pospelov:2007mp, ArkaniHamed:2008qn}}, in which the DM annihilates to two light vector bosons $V$ (or a similar mediator) each of which then decay to standard model particles and can be observed~\cite{Pospelov:2008jd, Rothstein:2009pm, Bell:2011sn}. If the decay is primarily to neutrinos, one gets two neutrino pairs in the final state. There is a recent proposal in \cite{Aarssen:2012fx}, which addresses some of the purported small-scale problems in $\Lambda$CDM, also the DM annihilation to neutrinos is enhanced and hence this model can be tested using neutrino telescopes.

The neutrino spectrum has a box-like structure 
\begin{equation}
\frac{dN_{\nu}}{dE_{\nu}}=\dfrac{4}{\Delta E}\Theta(E_{\nu}-E_-)\Theta(E_+-E_{\nu})\,.
\end{equation}
where $\Theta$ denotes the Heaviside-theta function. The maximum and minimum energy of the neutrino in this case is denoted by $E_{\pm}=(m_{\chi}\pm \sqrt{m_{\chi}^2-m_V^2})/2$. The width of the box function by $\Delta E=\sqrt{m_{\chi}^2-m_V^2}$. Neutrino oscillation ensures that the ratio of the neutrino fluxes reaching the neutrino detector is 1:1:1.

\section{Detection and Backgrounds}       \label{sec:NeuDet}

\subsection{Neutrino detection}
In a km$^3$-scale neutrino telescope like IceCube~\cite{IceCube:2011ab} or KM3NeT~\cite{KM3NeT}, neutrinos are detected as two different types of events: tracks and cascades.

\subsubsection{Tracks}The tracks are produced by the charged current interaction of the muon neutrinos and antineutrinos. At these high energies, the muons are produced by muon neutrinos interacting with the detector material or with the surrounding medium and the muon track is generally not contained inside the detector~\cite{Kistler:2006hp}. Due to the long range of the muon tracks, the effective volume of the detector is increased and the increase in the volume is determined by the range of the muon, of a given energy $E$, given by integrating the energy loss rate
\begin{eqnarray}
\label{eq:energy loss of the muon}
-\frac{dE}{dX}=\alpha+\beta E~,
\end{eqnarray}
where $X$ denotes the column density (in units of ${\rm g\,cm^{-2}}$). For our calculations, we take $\alpha=2$\,MeV\,cm$^2$\,g$^{-1}$ and $\beta=4.2\,\times 10^{-6}$\,cm$^2$\,g$^{-1}$~\cite{Kistler:2006hp}. 

The number of neutrinos detected per unit energy interval for muon tracks, which are not contained inside the detector is given by~\cite{Kistler:2006hp}
\begin{eqnarray}
\label{eq:though-going muon spectrum}
\frac{dN_{\mu}}{dE_{\mu}}\bigg|_{\rm{tracks}}&=&\frac{N_A ~\rho ~T ~A_{\rm{det}}}{\rho(\alpha+\beta E_{\mu})}  \nonumber\\
&\times& \int _{E_{\mu}}^{\infty}dE_{\nu} \frac{d\Phi_{\nu}}{dE_{\nu}}\sigma_{\rm{CC}}(E_{\nu}) e^{-\frac{L}{\lambda}} ~.
\end{eqnarray}
In the above formula, $N_A$ denotes the Avogadro's number, $\rho$ represents the density of the detector material, $\it{T}$ is the time of observation, $A_{\rm{det}}$ denotes the area of the detector, $\sigma_{\rm{CC}}(E_{\nu})$ denotes the charged current cross section of the muon neutrino with the detector material or its surroundings, $L$ the length traveled by the neutrino in the Earth, and $\lambda$ is the mean free path of the neutrino.

The factor $A_{\rm{det}}\rho^{-1}/(\alpha+\beta E_{\mu})$ accounts for the increased volume of the detector due to the long muon range. We take $A_{\rm{det}}$ = 1\,km$^2$ and $\it{T}$ = 10 years for the $\chi \chi \, \rightarrow \, \mu^+\mu^-$, $\chi \chi \, \rightarrow \, \nu \overline{\nu}$, $\chi \chi \, \rightarrow \, t \overline{t}$, and $\chi \chi \, \rightarrow \, \nu \overline{\nu} \nu \overline{\nu}$ channels. The values of $\sigma_{\rm{CC}}(E_{\nu})$ are taken from ~\cite{Gandhi:1998ri}. The exponential suppression is due to the absorption of very high energy neutrinos ($\gsim$\,100 TeV) as it passes through the Earth. The mean free path of the neutrinos in Earth matter is given by $\lambda=1/(n\,\sigma_{\rm{tot}})$, where $n$ denotes the number density of target particles and $\sigma_{\rm{tot}}$ denotes the total neutrino-nucleon cross section, which we take from ~\cite{Gandhi:1998ri}. For the energies considered here, the exponential factor is $\sim$ 1.

For non-contained muon track events in IceCube, the energy is obtained by using Eq.\,(\ref{eq:energy loss of the muon})  after measuring the muon energy loss inside the detector~\cite{Abbasi:2010ie}. The limits of the integral in Eq.\,(\ref{eq:though-going muon spectrum}) imply that a muon of energy $E_{\mu}$ can be produced by any $\nu_{\mu}$ with an energy $\geq$ $E_{\mu}$. 

The energy range that we explore using muon tracks is (100 GeV -- 100 TeV). Energy resolution for muon tracks is approximately 0.3 in units of log$_{10}E$~\cite{Resconi:2008fe}. Following~\cite{Mandal:2009yk}, we take the energy bin for signal calculation to be  $\left(\mbox{max}(E_{\mbox{\small{thres}}}, m_{\chi}/5), m_{\chi}\right)$. This energy bin is much bigger than the energy resolution of IceCube/KM3NeT~\cite{Abbasi:2010ie}. We expect a full spectral analysis by IceCube/KM3NeT collaboration to give much better sensitivity as the shape of the signal and background spectra are very different. In this regard, the results presented here can be treated as conservative.

Angular pointing for tracks is quite accurate. For neutrino energies $\gsim$\,100 GeV, the angular resolution is within 0.5$^{\circ}$ and 1$^{\circ}$~\cite{Abbasi:2010ie}.

\subsubsection{Cascades}
Charged current interactions of $\nu_e$ and $\nu_{\tau}$ and their antiparticles, and neutral current interactions of all flavors of neutrinos produce cascades. The electron produced due to the charged current interaction of the $\nu_e$ with the detector material causes an electromagnetic cascade in the detector. The $\tau$-lepton produced due to the charged current interaction of the $\nu_{\tau}$ with the detector material produces a hadronic cascade from its hadronic decay products and an electromagnetic cascade from the electrons arising from $\tau$ decay. The non-leptonic final states in a neutral current interaction causes a hadronic cascade in the neutrino telescope. These cascades are contained inside the detector, act as almost point sources of light, and are calorimetric. The cascade search also has lower atmospheric neutrino background~\cite{Beacom:2004jb}. Cascades has been detected in IceCube~\cite{Abbasi:2011ui} and recently also in DeepCore~\cite{Ha:2012ww}.

The number of neutrino events detected via cascades per unit energy interval is given by~\cite{Mandal:2009yk}
\begin{eqnarray}
\label{eq:cascade event}
\frac{dN_{\nu}}{dE_{\nu}}\bigg|_{\rm{casc}}&=&N_{A} ~T~V_{\rm{casc}} \nonumber\\
&\times& \bigg(\sigma_{\rm{CC}}(E_{\nu})\frac{d\Phi_{\nu_{e,\tau}}}{dE_{\nu}}
+\sigma_{\rm{NC}}(E_{\nu})\frac{d\Phi_{\nu_{e,\mu,\tau}}}{dE_{\nu}}\bigg)\,,{\phantom~~~~}
\end{eqnarray}
where $V_{\rm casc}=0.02\,\rm km^{3}$ is the volume available for cascades in a detector like IceCube-DeepCore and $\sigma_{\rm{NC}}$ the neutral current cross section of neutrinos, which we take from Ref.~\cite{Gandhi:1998ri}. Other symbols have meanings and values as previously defined.

The mass range of DM that we explore in the cascade analysis is (10 GeV -- 10 TeV).
The energy resolution for cascade like events is approximately 0.18 in units of log$_{10}E$~\cite{Resconi:2008fe}.
Following~\cite{Mandal:2009yk} we take the energy bin for signal calculation to be  $({\rm max}(E_{\rm thres}, m_{\chi}/5), m_{\chi})$. This energy bin is much larger than the energy resolution of IceCube/KM3NeT~\cite{Abbasi:2010ie}. We expect a full spectral analysis by IceCube/KM3NeT collaboration to give much better sensitivity as the shape of the signal and background spectra are very different. In this regard the results presented here can be treated as conservative.

Achieved angular resolution of cascades in IceCube is about 50$^\circ$, but is expected to be significantly improved in the future with more advanced reconstruction algorithms in DeepCore~\cite{Middell}. With a large angular resolution the background due to atmospheric neutrinos is overwhelming, and improving the resolution drastically cuts down background. Encouragingly, Auer~\cite{Auer:2009zz} discusses a procedure which can be used to reconstruct the angular resolution of cascades to about 5$^{\circ}$ in KM3NeT. We shall show that with such improved angular resolution, the sensitivity to DM annihilation cross section by cascades increases significantly.

\subsection{Detector Configurations and Backgrounds}

While calculating the sensitivity to the DM annihilation cross section using muon tracks, we assume that the neutrino telescope only looks at upgoing tracks. This means that IceCube will look at galaxy clusters in the northern hemisphere and KM3NeT, while using muon tracks for their analysis, will look at galaxy clusters in the southern hemisphere. Looking for upgoing tracks eliminates the background caused by downgoing atmospheric muons.

For the cascade analysis we shall assume that KM3NeT includes a DeepCore-like low energy extension, which we call KM3NeT-Core. We assume the mass of the KM3NeT-Core to be the same as that of DeepCore. Similar to DeepCore, we shall assume that KM3NeT-Core will use the remainder of the KM3NeT as a veto. Such an arrangement allows the low energy extension in KM3NeT to have a 4$\pi$ field of view, and therefore this low energy extension in KM3NeT can also be used to detect galaxy clusters in the northern hemisphere. 

With the configurations explained in the previous two paragraphs, the backgrounds in both track and cascade analyses are due to atmospheric neutrinos and neutrinos from cosmic ray interactions in the galaxy cluster.

The measured atmospheric $\nu_{\mu}$, $\overline{\nu}_{\mu}$ flux for $E_{\nu}$ in the range of (100 GeV -- 400 TeV) is reported in~\cite{Abbasi:2010ie}. The measured spectrum is fit well by the angle-averaged atmospheric neutrino spectrum given in~\cite{Erkoca:2010vk}:
\begin{eqnarray}
\label{eq:angle averaged atmospheric neutrino spectrum}
\dfrac{d\phi_{\rm{atm.}}}{dE_{\nu}}&=&{\Phi^0_{\rm atm}}  E_{\nu}^{-2.74}\times 10^{17}\,{\rm GeV^{-1}km^{-2}yr^{-1}}\\
&\times& \left(\dfrac{{\rm ln}\,(1+0.024\,E_{\nu})}{1.33\,E_{\nu}}
+ \dfrac{{\rm ln}\,(1+0.00139\,E_{\nu})}{0.201\,E_{\nu}}\right)\,,\nonumber 
\end{eqnarray}
where $\Phi^0_{\rm atm}=1.95 $ for neutrinos and 1.35 for antineutrinos, and $E_{\nu}$ is in GeV. The atmospheric $\nu_e$ flux is taken from~\cite{Honda:2006qj} and the $\nu_{\tau}$ flux is from~\cite{Pasquali:1998xf}.  

In addition to the atmospheric neutrino, the neutrinos produced by cosmic ray interactions inside the galaxy cluster also acts as an additional background. We take the neutrino flux produced in cosmic ray interaction in the galaxy clusters from the calculations by Murase~${et}$~$al.$~\cite{Murase:2008yt}. They consider acceleration of cosmic rays with energies between $10^{17.5}\,\rm eV$ and $10^{18.5}\,\rm eV$ in shocks in galaxy clusters. In a $1^\circ\times1^\circ$ angular bin, they estimate $\lsim$1 ($\nu_{\mu}+\overline{\nu}_{\mu}$) event per year above 1 TeV. Although this estimate is somewhat model-dependent, we emphasize that this is an essential background that one has to take into account while searching for neutrinos from DM annihilation in galaxy clusters. If it turns out that galaxy clusters are not sources of cosmic rays in this energy range then this background can be lower but we assume the larger background rate for conservative estimates.

\section{Results}       \label{sec:Results}
In this section we calculate the neutrino fluxes observed for the four annihilation channels chosen above, and compare them with the expected backgrounds to determine the sensitivity in the $\langle \sigma v \rangle$-$m_{\chi}$ plane for each channel. However, before we proceed to results specific to each channel, we identify some broad features.

The first key result is regarding the optimal size of the region of interest. The signal we are looking for is proportional to $\int_0^{\Delta \Omega} \left(j_{\rm{sub}}+j_{\rm{NFW}}\right) d\Omega$. We scale this quantity with the local DM density squared and the distance to the Galactic centre from the Sun, as in Eq.\,(\ref{eq:scaled J}), to get 
\begin{eqnarray}
\label{eq:Jtildetot}
\tilde{J}_{\rm{tot}}(\psi _{\rm max})= \int_0^{\Delta\Omega}\left(\tilde{j}_{\rm{sub}}+\tilde{j}_{\rm{NFW}}\right) d\Omega\,,
\end{eqnarray}
where $\Delta \Omega$ depends on the angular radius $\psi_{\rm max}$ of the region of interest. The fluctuations in the atmospheric neutrino background are proportional to $\sqrt{\Delta \Omega(\psi_{\rm max})}$. Therefore, the signal-to-noise ratio is approximately proportional to $\tilde{J} _{\rm{tot}}(\psi_{\rm max})/\sqrt{\Delta \Omega(\psi_{\rm max})}$.

\begin{figure}[t]
\includegraphics[width=1 \columnwidth]{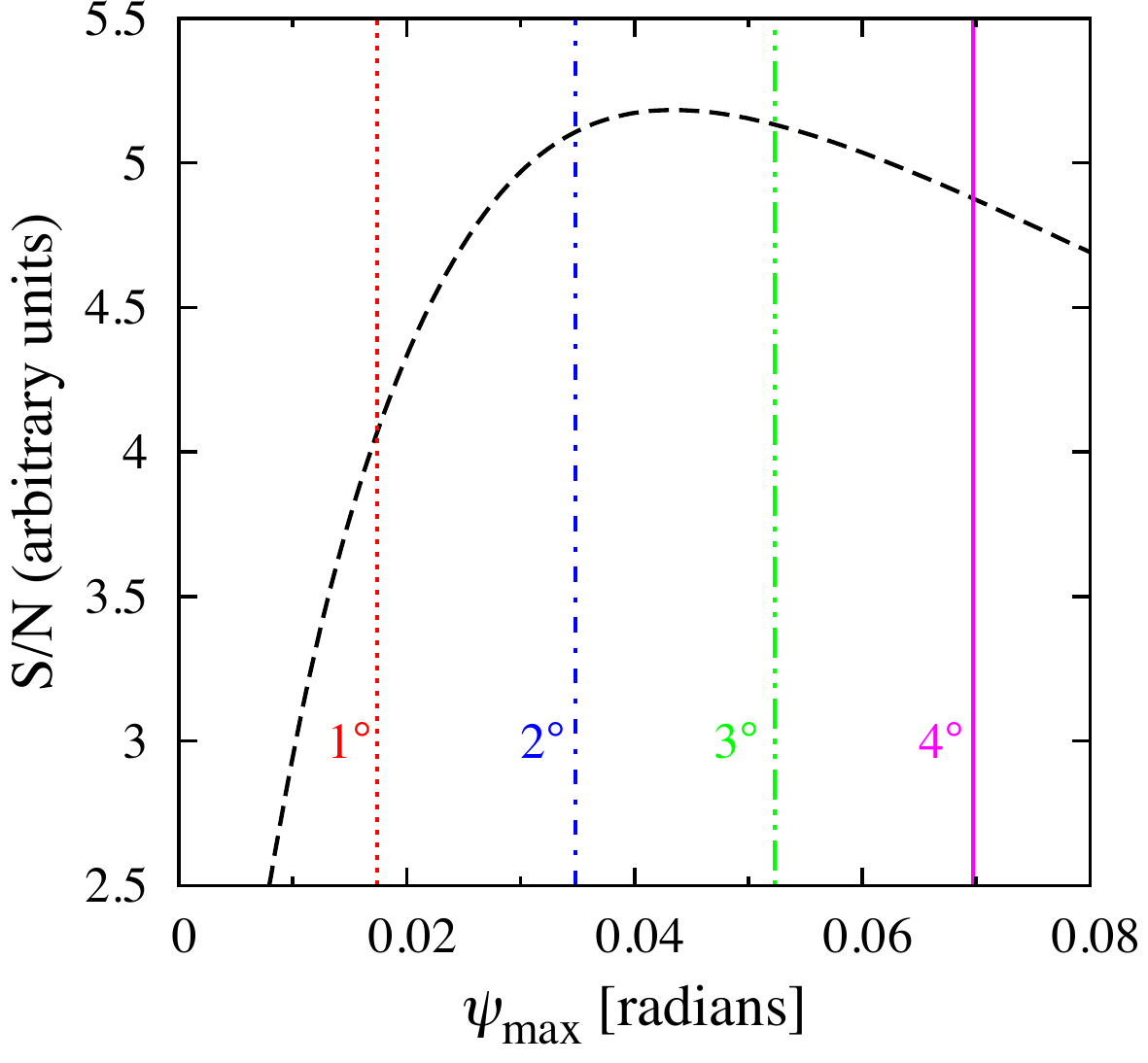}
\caption{Ratio of approximate signal-to-noise versus the angular size of the chosen region of interest around the Virgo galaxy cluster. The vertical lines show the values of some representative angular radii in degrees.}
\label{fig:J over square root delta omega}
\end{figure}

\begin{figure*}[!htbp]
\centering
\includegraphics[angle=0.0, width=0.45\textwidth]{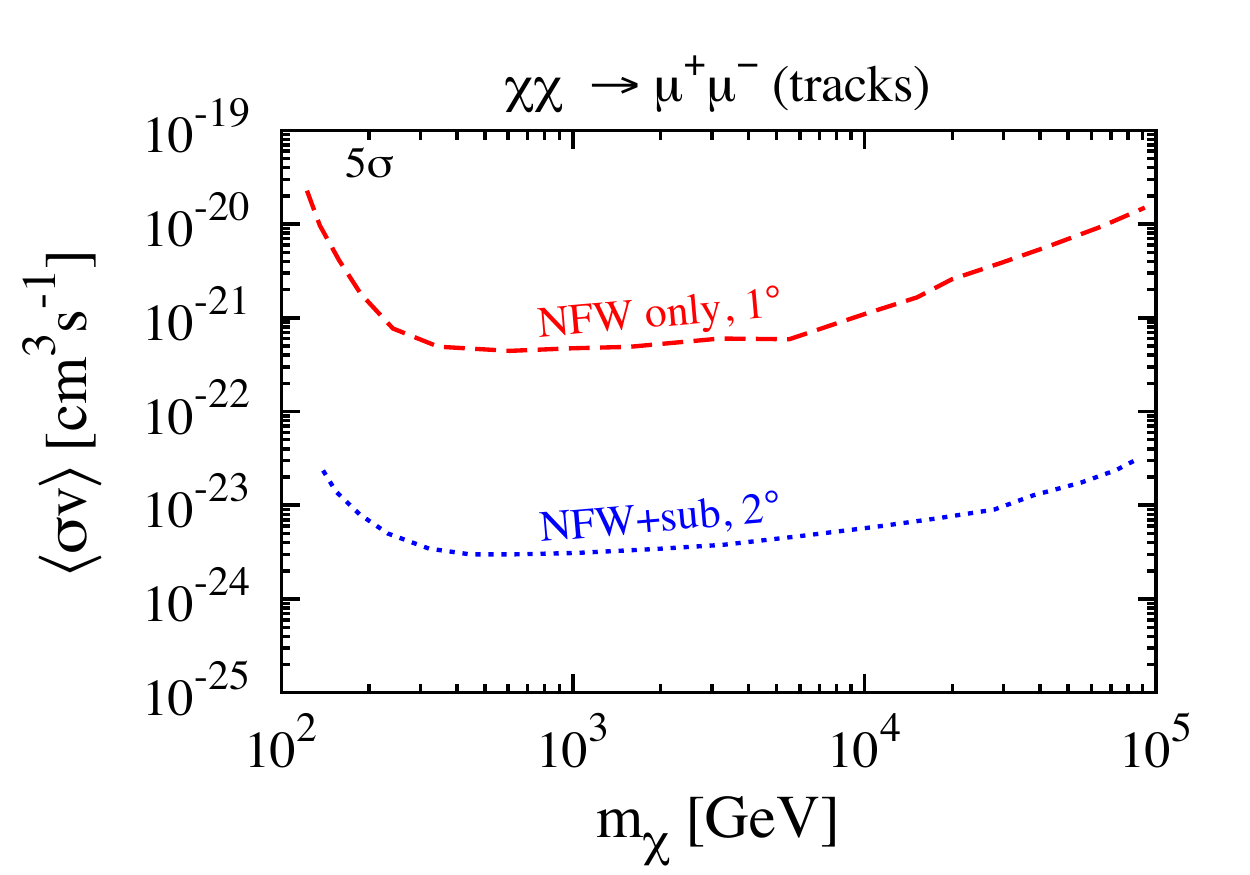}
\includegraphics[angle=0.0, width=0.45\textwidth]{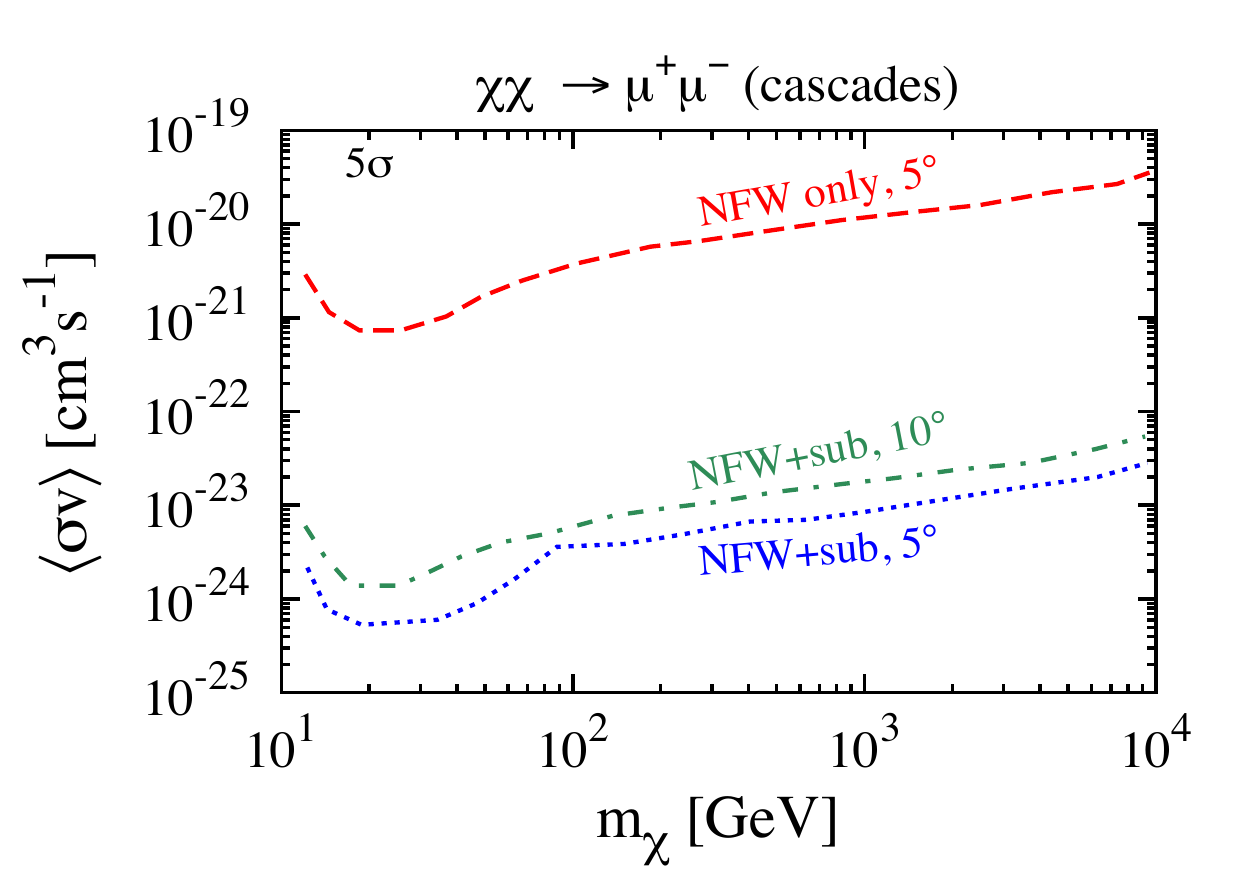}
\includegraphics[angle=0.0,width=0.45\textwidth]{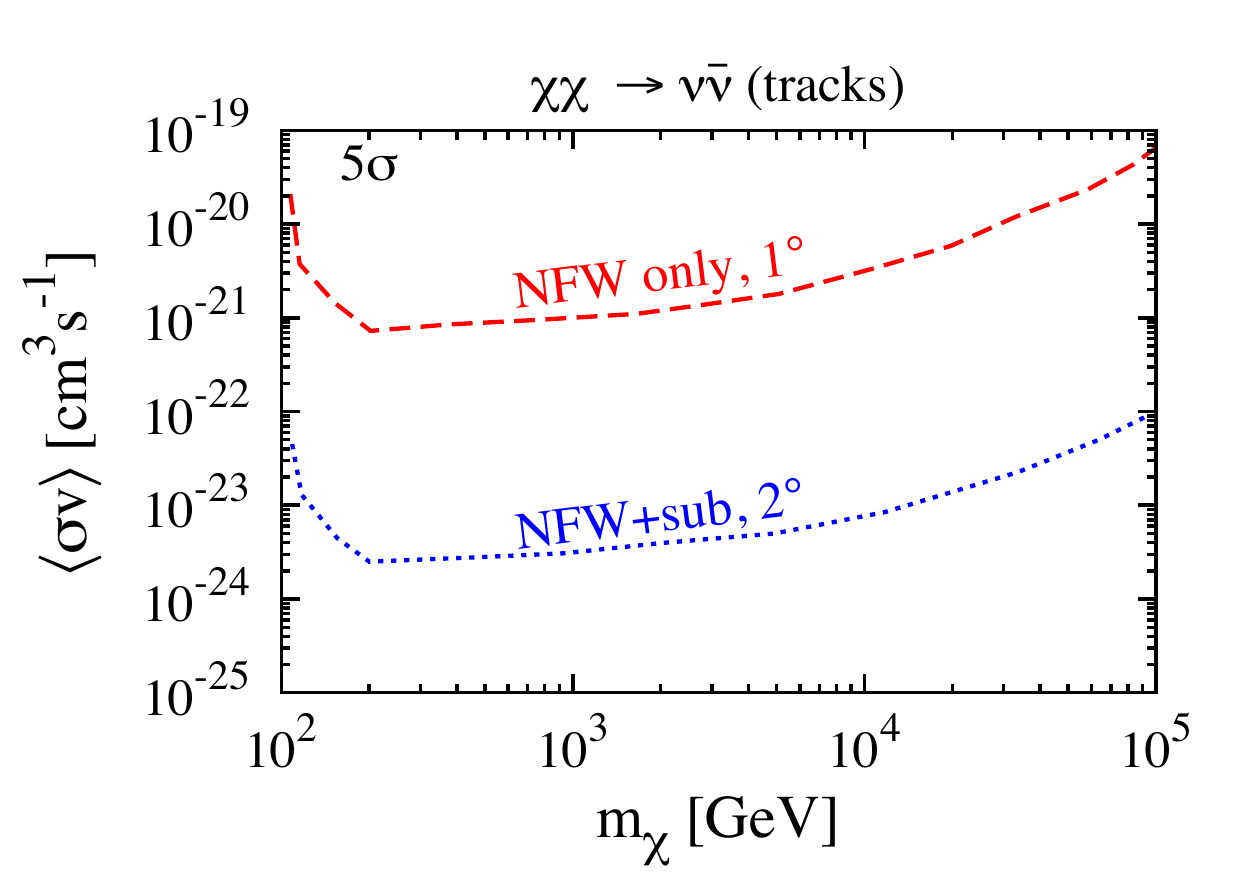}
\includegraphics[angle=0.0,width=0.45\textwidth]{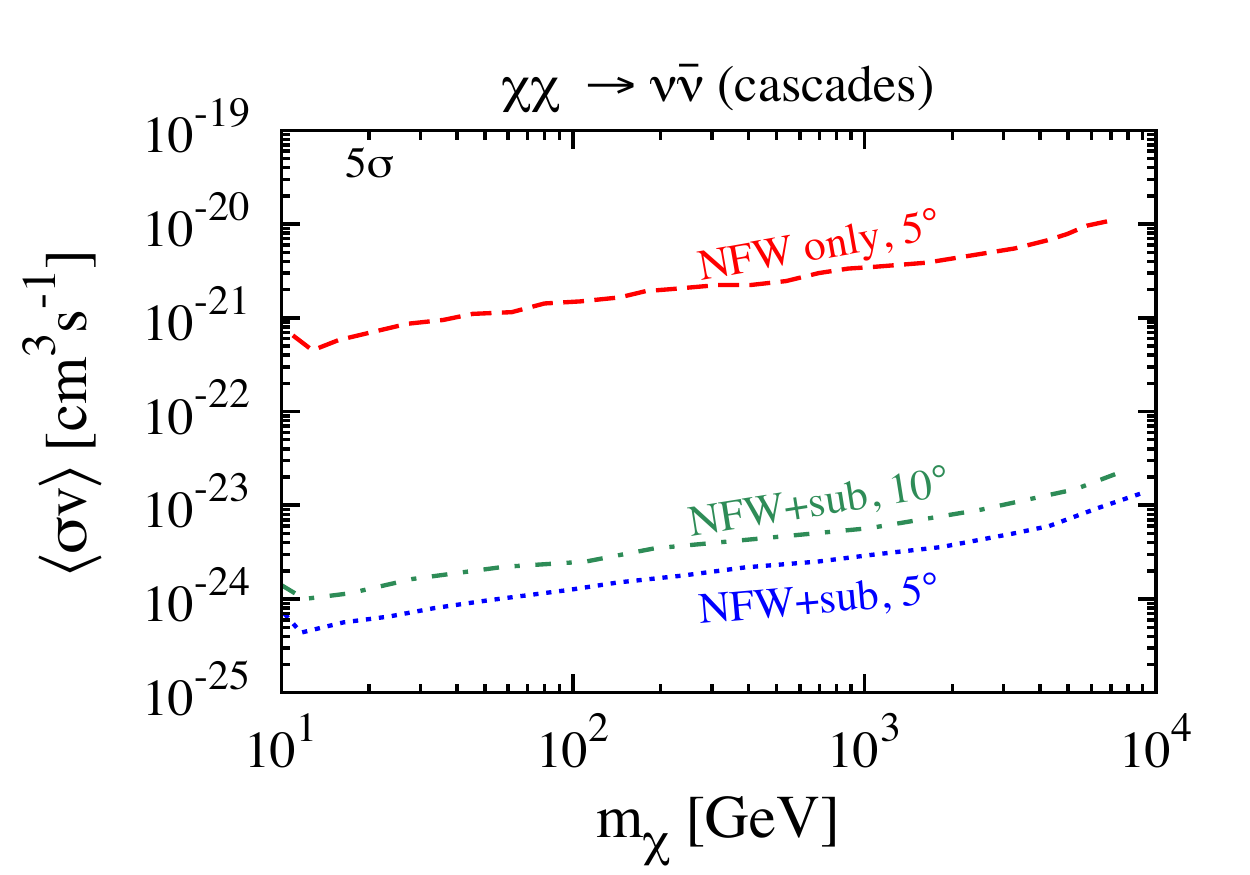}
\includegraphics[angle=0.0,width=0.45\textwidth]{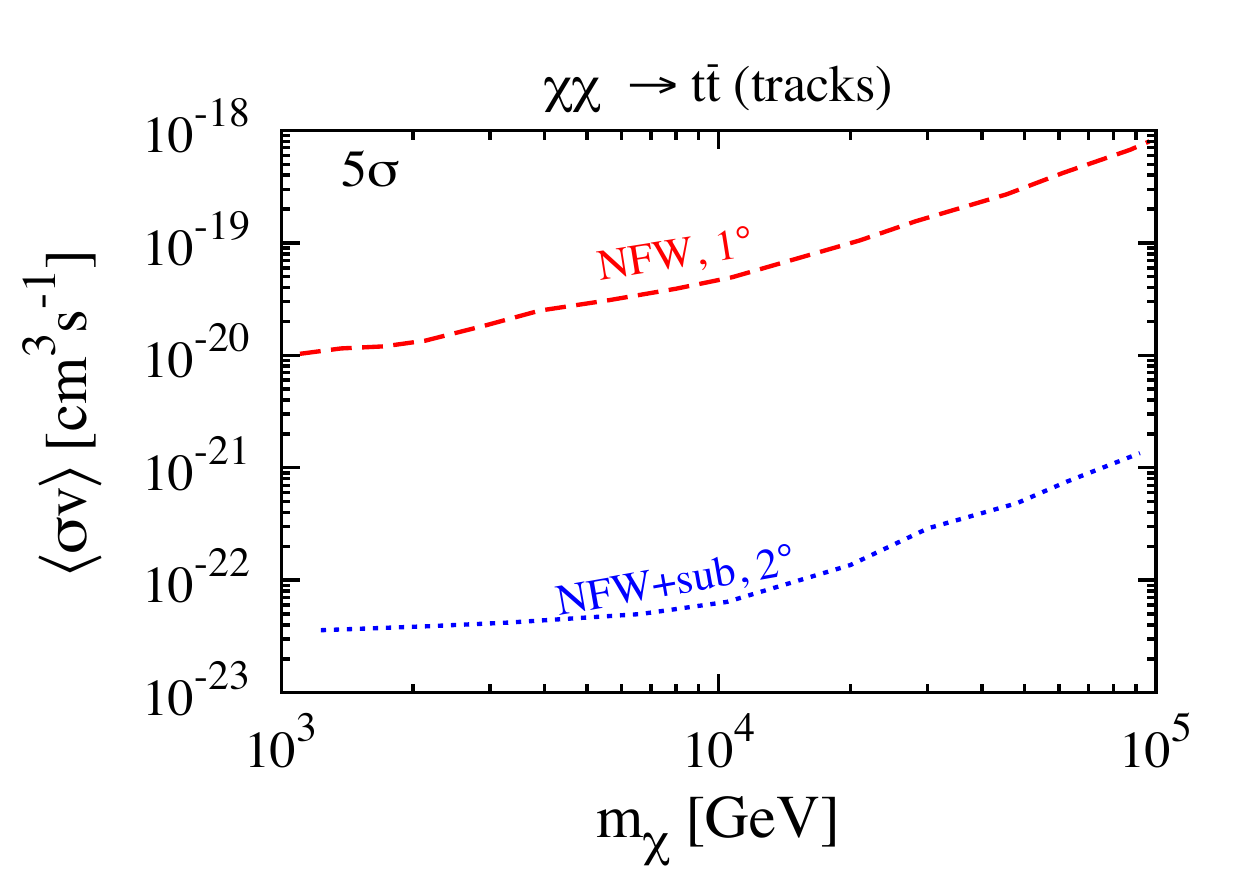}
\includegraphics[angle=0.0,width=0.45\textwidth]{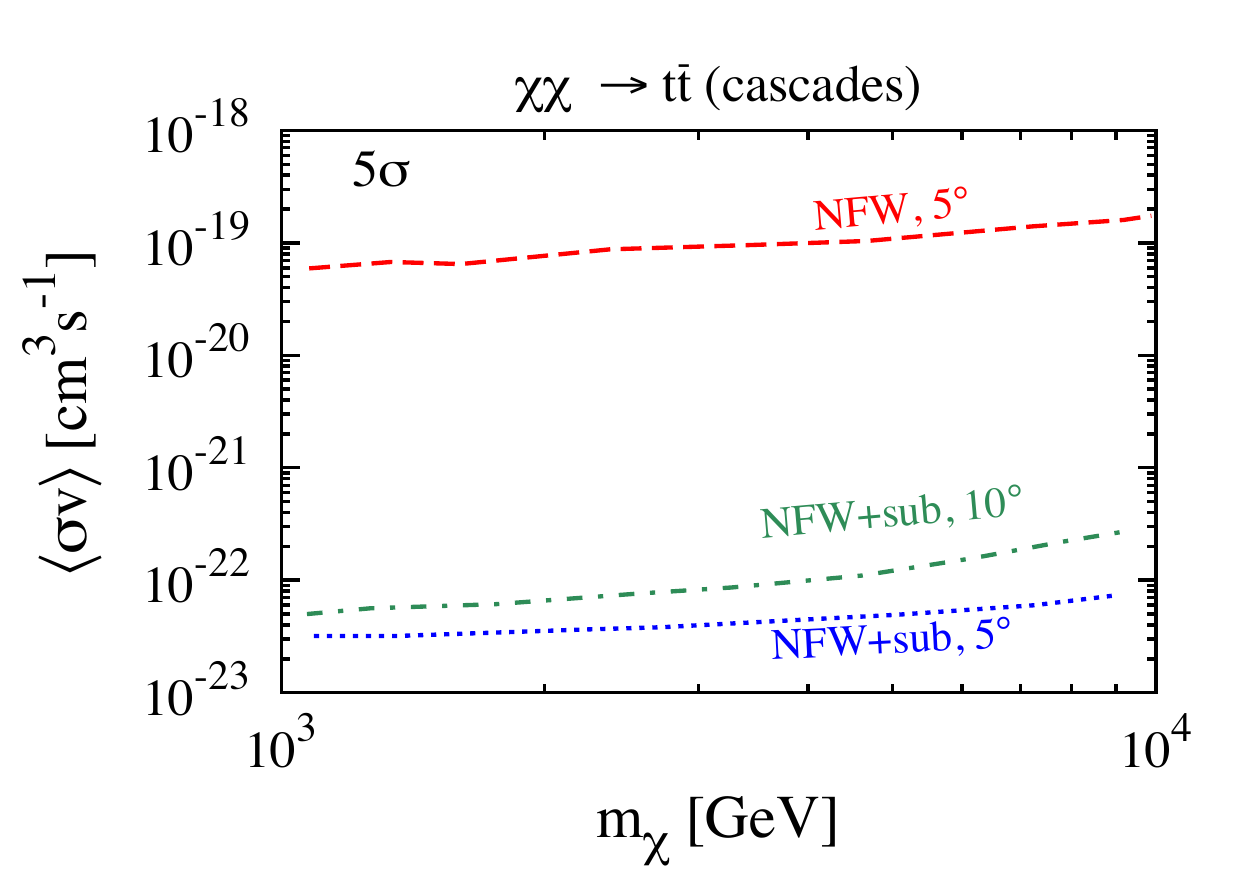}
\includegraphics[angle=0.0,width=0.45\textwidth]{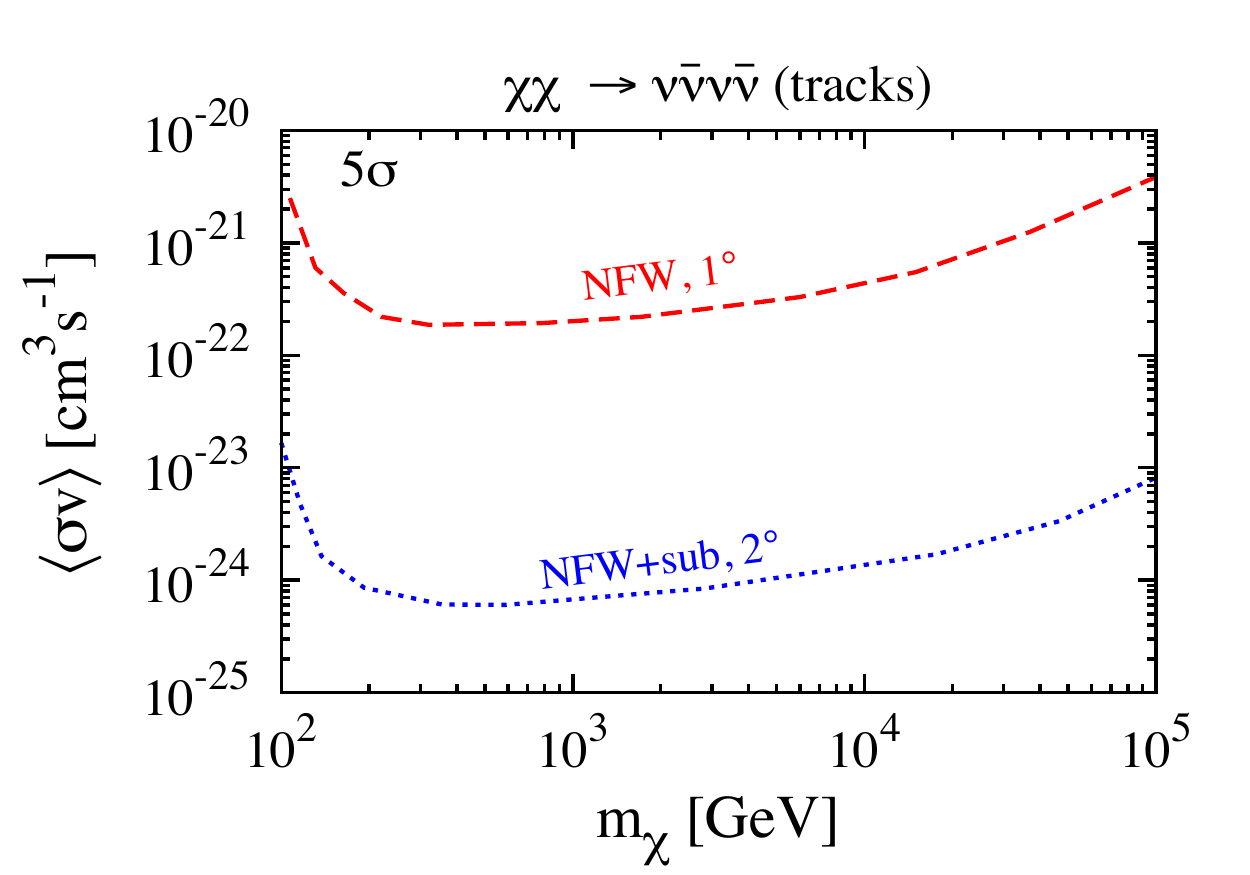}
\includegraphics[angle=0.0,width=0.45\textwidth]{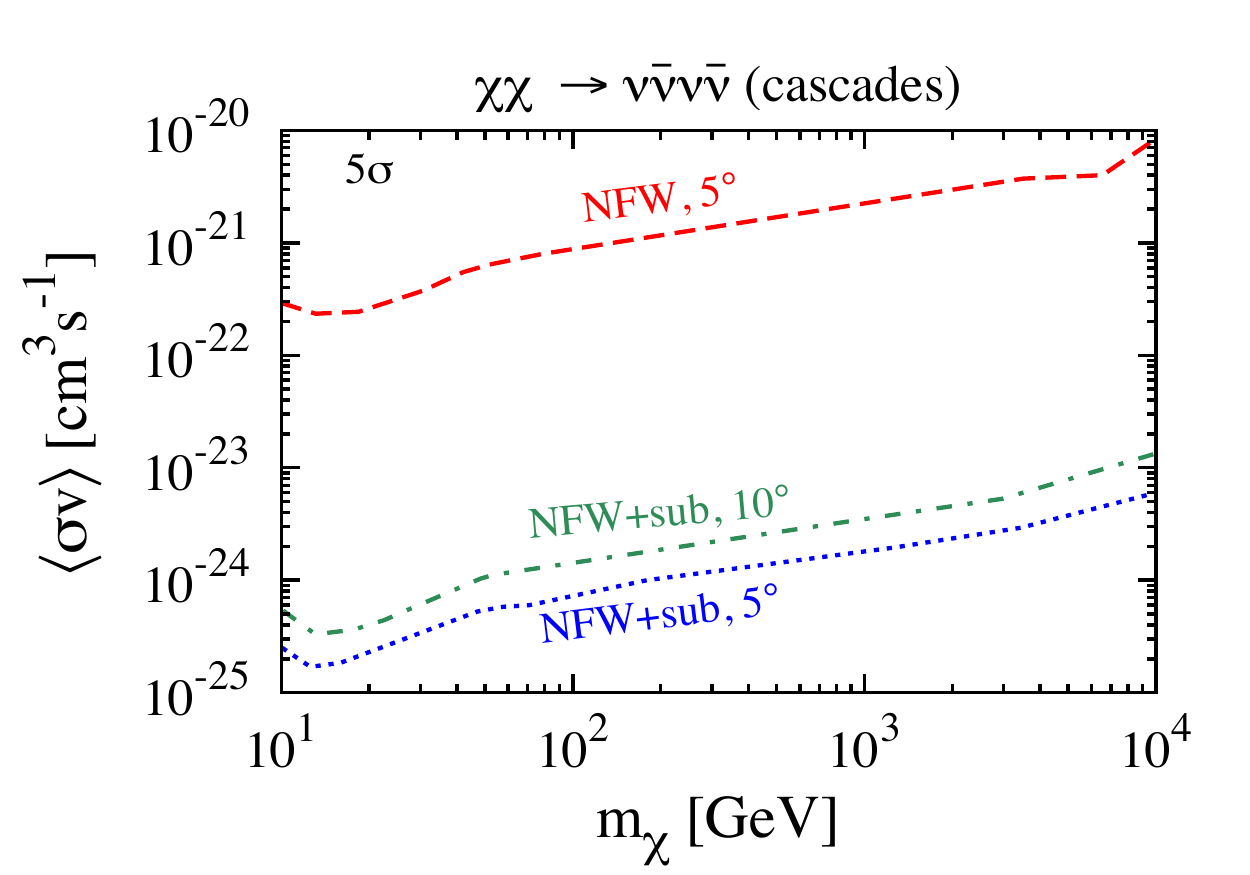}
\caption{Sensitivity to $\langle \sigma v \rangle$ versus the DM mass, for the annihilation channels $\chi \chi \rightarrow \mu^+ \mu^-$ (top panels),  $\chi \chi \rightarrow \nu \overline{\nu}$ (second panel from top), $\chi \chi \rightarrow t\overline{t}$ (third panel from top), and $\chi \chi \rightarrow \nu \overline{\nu} \nu \overline{\nu}$ (bottom panel) using 10 years of IceCube/KM3NeT-Core data for all the channels. Left panels: Sensitivity that can be obtained from muon tracks. Right panels: Sensitivity that can be obtained using cascades. The lines are labeled with the assumed DM density profile: smooth component only (NFW) or including both smooth and substructure components (NFW+sub), as well as the angular radius chosen for the region of interest ($1^{\circ},\,2^{\circ},\,5^{\circ},$ and $10^{\circ}$). We use the DM substructure modeling of~\cite{Han:2012au}. A more conservative DM substructure modeling following~\cite{SanchezConde:2011ap} will worsen the sensitivity in each case by a factor of $\sim$20.}
\label{fig:sensitivity}
\end{figure*}

In Fig.\,\ref{fig:J over square root delta omega} we show this ratio for the Virgo galaxy cluster, as a function of the chosen angular radius $\psi_{\rm max}$ of the region of interest. We can see that due to the extended nature of the DM substructure profile, a region with $\approx2^{\circ}$ angular radius around the galaxy cluster gives the best signal-to-noise ratio. We have verified using our numerical results that the sensitivity obtained with a 2$^\circ$ observation window is about a factor of 1.5 better using a 1$^\circ$ window.

Neutrino telescopes should therefore carefully optimize for the observation window. Selecting a circular region of diameter $\sim$~4$^{\circ}$ around the centre of the galaxy cluster and accepting signal events coming from that circular region appears to give the best signal-to-noise ratio. Depending on the specific DM profile of a galaxy cluster, this choice of angle may change but, in general, we expect that, for any nearby galaxy cluster the best signal-to-noise ratio is achieved by treating it as an extended source, as opposed to a point source.

The second key result is that the presence of substructures gives three orders of magnitude more promising results than the smooth NFW profile alone. This boost provided by the substructures make galaxy clusters an exciting target for neutrino telescopes. Using track-like events, the sensitivity is typically in the range $\sigv\gsim(10^{-24} - 10^{-22})\,{\rm cm^3\,s^{-1}}$ at DM masses ($100\,\rm{GeV}-100\,\rm{TeV})$. At lower masses, the sensitivity worsens quickly because the events are below threshold (note the upturn in Figs.\,\ref{fig:sensitivity} and \ref{fig:comparison}). However the sensitivity also worsens at extremely high masses because as $m_\chi$ increases, the number of DM particles decrease for a given DM density, reducing the annihilation fluxes. 

The third key result we find is that if KM3NeT can reconstruct cascades with an angular resolution of $\sim$~5$^\circ$ and has a DeepCore-like low energy extension, a new window of observation opens up at lower DM masses ($10\,{\rm GeV}-100\,{\rm GeV}$). The sensitivity of neutrino cascade observations remain competitive with track analyses at masses up to $10\,{\rm TeV}$. This complementary measurement of muon tracks and cascades may be useful to explicitly determine the neutrino flavors in the DM annihilation products. We believe this should encourage the KM3NeT collaboration to improve their cascade pointing to $\lesssim5^\circ$ and include a DeepCore-like low energy extension in KM3NeT.

We emphasize that if neutrino telescopes detect a DM annihilation signal from galaxy clusters at a sensitivity forecasted in this paper and if the emission profile is found to be extended, it will be a strong indication for the presence of substructures. If the neutrino signal is not extended but the cross section is comparable to what is forecasted to be testable then it will favor an enhanced annihilation cross section, rather than the presence of substructures.  A particle physics explanation of the enhanced DM annihilation cross section will then be required~\cite{ArkaniHamed:2008qn, Feldman:2008xs, Ibe:2008ye}. If a signal is not detected at an annihilation cross section testable at neutrino telescopes, then it will either constrain the minimum DM substructure mass and abundance or the annihilation cross section. In that situation, one will have to first infer the DM annihilation cross section from some other astrophysical source to infer something about the minimum DM substructure mass and the DM substructure distribution.

\subsection{$\chi \chi ~\rightarrow ~\mu^+\mu^-$ }

In Fig.\,\ref{fig:sensitivity} (top left panel), we show the sensitivity to $\sigv$ for the DM annihilation to $\mu^+\mu^-$, that can be achieved by observing muon tracks at IceCube. The sensitivity is maximum at $m_{\chi}\approx 500$\,GeV, where cross sections as small as $\sigv \approx10^{-24}{\rm cm^3s^{-1}}$ may be probed by IceCube. By observing the Milky Way halo, IceCube has already constrained the value of this annihilation cross section to be $\sigv \lsim 10^{-22}\,{\rm cm^3s^{-1}}$ for a DM mass of about 1 TeV~\cite{Abbasi:2011eq}. We expect that the sensitivity obtained from observing the Virgo galaxy cluster will improve the above limit by about one order of magnitude if no detection is obtained.

Han \etal\,\cite{Han:2012au}, recently found evidence of extended gamma ray emission from the Virgo cluster, and the limit on the annihilation cross section that they obtained is $\sigv\approx\,10^{-25}\,{\rm cm^3s^{-1}}$. Although, in principle, this channel is observable at IceCube, we find that IceCube does not have the sensitivity needed to test this claim. Note that after the publication of the first version of this paper, it was found by several groups that the extended gamma ray emission in the Virgo cluster is due to the presence of new gamma-ray sources and not due to DM annihilation~\cite{MaciasRamirez:2012mk,Han:2012uw}.

The sensitivity that can be obtained by observing cascades is shown in Fig.\,\ref{fig:sensitivity} (top right panel). As KM3NeT is still under development, we show the constraints using two plausible choices for its angular resolution. We find that the sensitivity obtained from cascades is almost comparable with that obtained from muon tracks. The best sensitivity is achieved around a DM mass of around 20\,GeV where a sensitivity to $\sigv\approx \mathcal{O}(10^{-25}\,{\rm cm^3s^{-1}})$ is reached, representing an order of magnitude improvement over the best sensitivity obtainable by observing tracks.

\subsection{$\chi \chi ~\rightarrow ~\nu \overline{\nu}$}
In Fig.\,\ref{fig:sensitivity} (second from top and left panel), we show the expected sensitivity to self annihilation cross section $\sigv$ for $\chi \chi ~\rightarrow ~\nu \overline{\nu}$, for $m_{\chi}$ in the range of ($100\,{\rm GeV}-100\,{\rm TeV}$), by detecting track-like events. The sensitivity is strongest at $m_{\chi}\approx~500$\,GeV where the annihilation cross sections larger than $\sigv\approx \mathcal{O}(10^{-24}{\rm cm^3s^{-1}})$ can be probed. Due to the presence of substructures, we again get a sensitivity which is stronger by about one order of magnitude than the constraint obtained by IceCube when looking for this annihilation signal at the Milky Way Galactic halo~\cite{Abbasi:2011eq}. Since the spectra of the signal and background are very different in this case, we expect that a much better sensitivity can be achieved due to a spectral analysis by the IceCube collaboration for the same exposure.

We now consider cascade signals from this annihilation channel, in a DeepCore-like low energy extension in KM3NeT. For DM masses between 30\,GeV and 10\,TeV, the projected sensitivity is shown in Fig.\,\ref{fig:sensitivity} (second from top and right panel). In the low DM mass range, sensitivity to annihilation cross sections $\sigv\approx\mathcal{O}(10^{-25}$\,cm$^3$s$^{-1})$ can be reached. As can be seen from the plot, KM3NeT will have a unique opportunity to probe this part of the parameter space if it employs a low energy extension.

\subsection{$\chi \chi ~\rightarrow ~t \overline{t}$}
We now look at the sensitivity that can be obtained to the $\sigv$ for DM annihilation via {$\chi \chi \rightarrow t \overline{t}$, by detecting track-like events. We show the sensitivity to $\sigv$, for $m_{\chi}$ in the range ($1\,{\rm TeV}-100\,{\rm TeV}$), in Fig.\,\ref{fig:sensitivity} (third from top and left panel). Annihilation cross sections  larger than $\sigv \approx \mathcal{O}(10^{-23}{\rm cm^3s^{-1}})$ can be probed for DM mass in the range ($1\,{\rm TeV}-10\,{\rm TeV}$). The constraints weaken for DM masses heavier than 10 TeV.

Observation of cascades in a DeepCore-like low energy extension in KM3NeT give similar sensitivity in the $(1\,{\rm TeV}-10\,{\rm TeV})$ mass range. The expected sensitivity is shown in Fig.\,\ref{fig:sensitivity} (third from top and right panel). Annihilation cross sections $\langle \sigma v \rangle\approx \mathcal{O}(10^{-23}{\rm cm^3s^{-1}})$ may be probed using the Virgo galaxy cluster.

\subsection{$\chi \chi ~\rightarrow VV \rightarrow ~\nu \overline{\nu} \nu \overline{\nu}$}
In Fig.\,\ref{fig:sensitivity} (bottom left panel), we show the expected sensitivity to $\sigv$ for $\chi \chi ~\rightarrow ~\nu \overline{\nu} \nu \overline{\nu}$, for $m_{\chi}$ in the range of ($100\,{\rm GeV}-100\,{\rm TeV}$), by detecting track-like events. The strongest sensitivity is achieved at $m_{\chi}\approx 500$\,GeV, where the annihilation cross sections larger than $\sigv\approx \mathcal{O}(10^{-25}{\rm cm^3s^{-1}})$ can be probed. Due to the presence of substructures, we again get a sensitivity which is stronger by about three orders of magnitude than the constraint obtained when assuming only an NFW profile.

We now consider cascade signals from this annihilation channel, in a DeepCore-like low energy extension in KM3NeT. For DM masses between 10\,GeV and 10\,TeV, the projected sensitivity is shown in Fig.\,\ref{fig:sensitivity} (bottom right panel). In the low DM mass range, sensitivity to annihilation cross sections $\sigv\approx \mathcal{O}(10^{-25}$ cm$^3$s$^{-1})$ can be reached. KM3NeT will have a unique opportunity to probe this part of the parameter space, which is not accessible by tracks, if it employs a low energy extension.

Neutrino telescopes have not searched for neutrinos from the annihilation channel $\chi\chi\,\rightarrow\,\nu\overline{\nu}\nu \overline{\nu}$, but as we show in Fig.\,\ref{fig:sensitivity} (bottom panels), the constraints obtained in this channel can be quite promising. In particular~\cite{Aarssen:2012fx} predicts enhanced emission in neutrinos, with $\sigv\sim10^{-24}{\rm cm^3s^{-1}}$ in galaxies. The velocity dependence of the cross section in this model will reduce the cross section in galaxy clusters, but we believe that, besides the Milky Way and dwarf galaxies, galaxy clusters may also offer an important test for this model due to the strong substructure enhancement.

\begin{figure*}[!thpb]
\centering
\includegraphics[angle=0.0,width=0.49\textwidth]{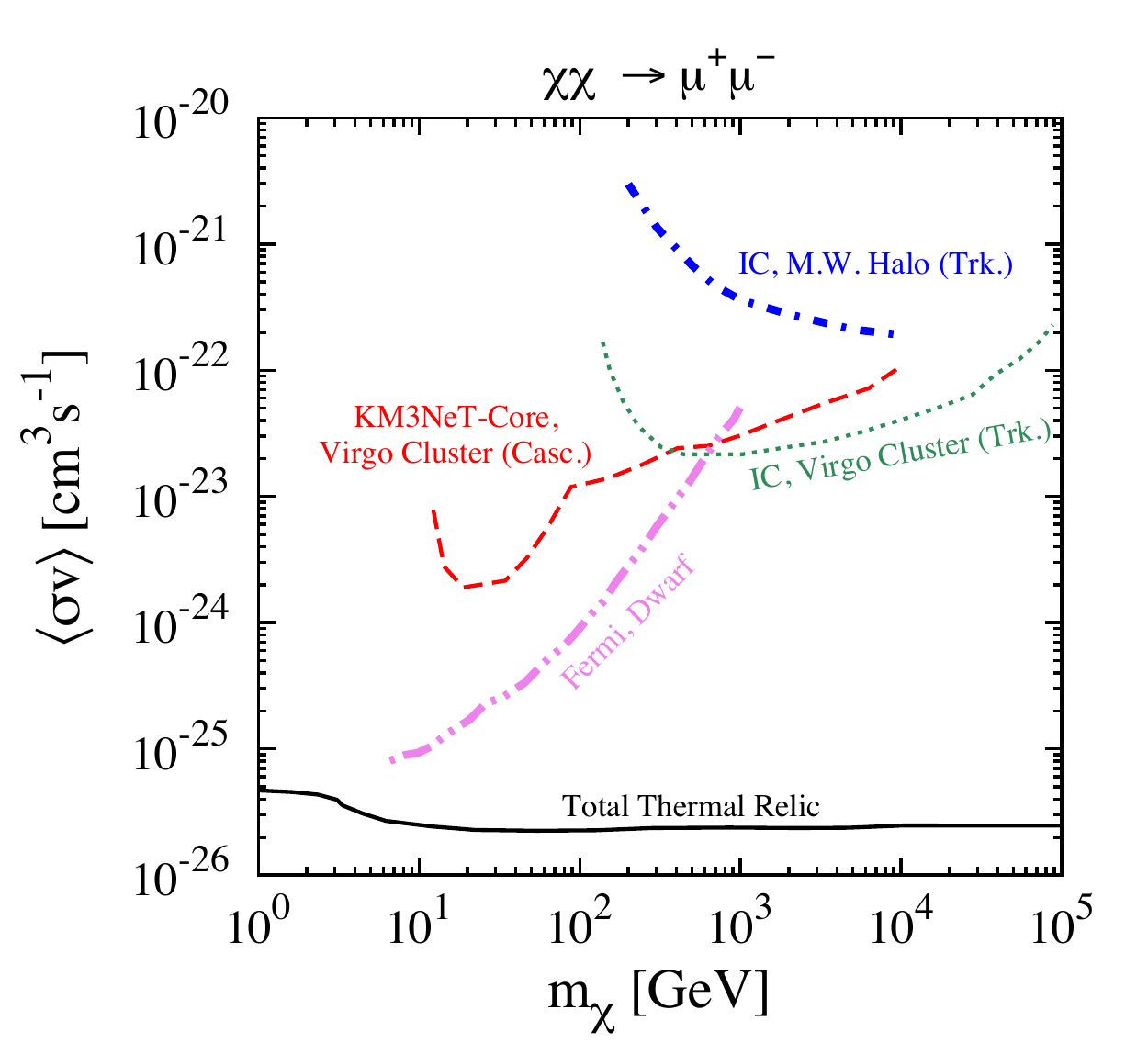}
\includegraphics[angle=0.0,width=0.49\textwidth]{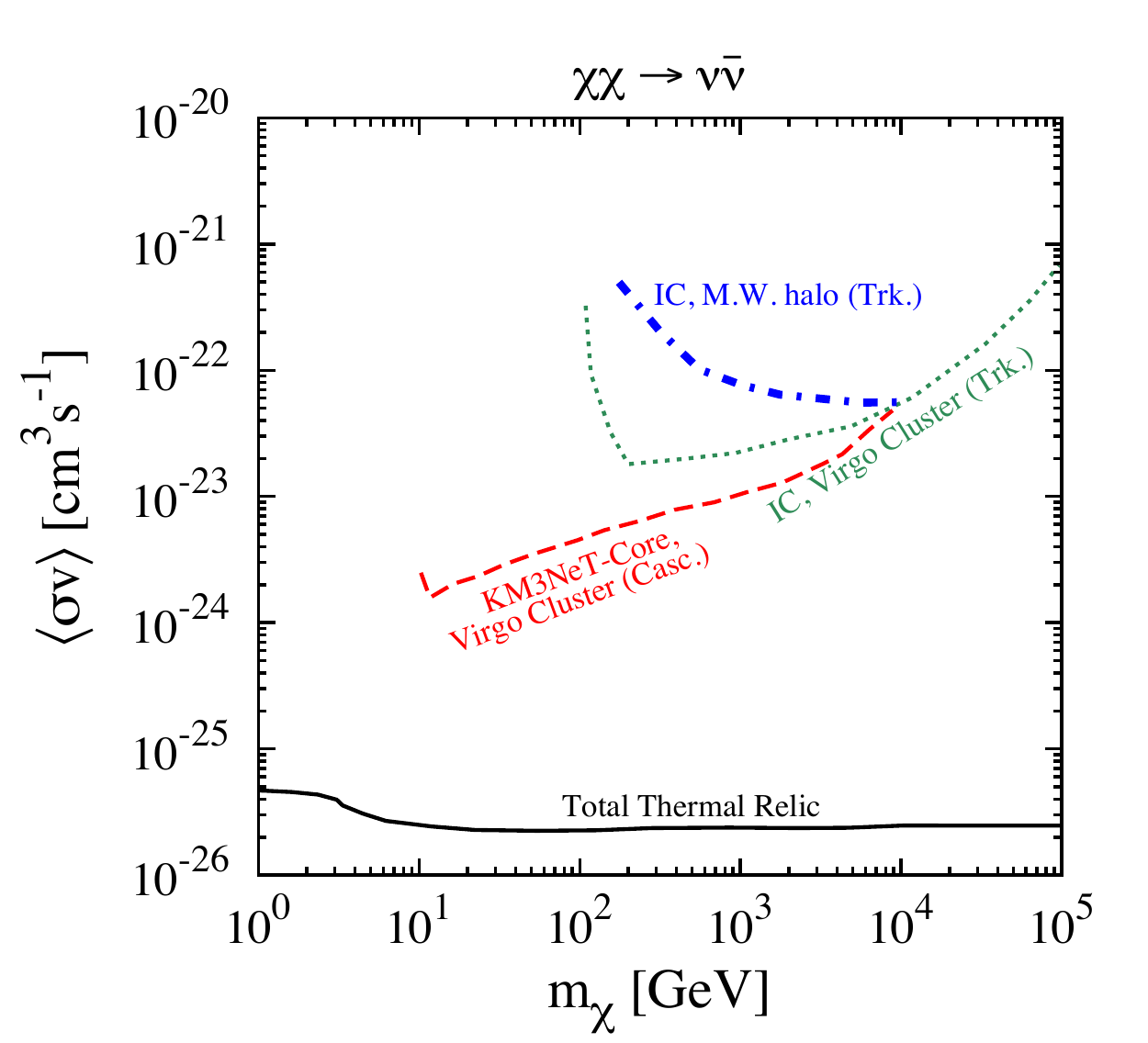}\\
\caption{Comparison of sensitivities of various experiments to DM annihilation in specific channels.  {\bf Left panel:} Annihilation via $\chi \chi\,\rightarrow\,\mu^+\mu^-$. {\bf Right panel:} Annihilation via $\chi \chi\,\rightarrow\,\nu\overline{\nu}$. The limit from Fermi-LAT analysis of gamma rays from dwarf galaxies~(\cite{Ackermann:2011wa}) is shown by the violet dash-dot-dot line. The limit obtained by the IceCube collaboration from observing the Milky Way galactic halo~(\cite{Abbasi:2011eq}) is shown by the blue dash-dot line. The sensitivity that is forecasted in this work by observing the Virgo galaxy cluster, for the same livetime as in~(\cite{Abbasi:2011eq}), is shown by the red dashed line (detecting cascades in low energy threshold KM3NeT-Core with a 5$^\circ$ angular resolution) and the green dotted line (detecting tracks in IceCube within a 2$^\circ$ angular radius). For comparison, we also show the total thermal relic annihilation cross section (Total Thermal Relic), as calculated by Steigman \etal~\cite{Steigman:2012nb}, by the black solid line.}
\label{fig:comparison}
\end{figure*}

\subsection{Comparison with limits from other experiments}

In this section, we compare the sensitivity that IceCube and KM3NeT can obtain by observing clusters of galaxies with limits obtained from other experiments. The main annihilation products observable in an indirect DM detection experiment are neutrinos and gamma rays. Both of these signals are not deflected by galactic or extragalactic magnetic fields and come directly into the detector from the source. For the muon track signal, the IceCube results are for 276 days of live time and in the 22 string configuration. To compare our calculations with the same exposure in~\cite{Abbasi:2011eq}, we use 276 days as the time of observation in Eqs.~(\ref{eq:though-going muon spectrum}) and~(\ref{eq:cascade event}) and take 1/4th of the number of neutrino events to mimic the 22-string detector. For the cascade signal, we show our results for a 276 days of livetime of the detector and we take the volume available for cascades as 0.02\,km$^3$.

We first show the various constraints on the $\chi\chi\,\rightarrow\,\mu^+\mu^-$ channel on the left panel in Fig.\,\ref{fig:comparison}. This channel can be detected by both neutrinos and gamma rays. The decay of the muons produce electrons which can produce gamma rays via inverse Compton and bremsstrahlung. These energetic electrons can also produce synchrotron radiation in the galactic and extragalactic magnetic fields but the synchrotron radiation is at a lower frequency. The neutrinos are produced in this channel via the decay of the muon.
 
At low DM masses ($m_{\chi}\lsim 100$ GeV), the constraints from Fermi-LAT using dwarf spheroidal galaxies are the most stringent~\cite{Ackermann:2011wa}. This limit weakens for DM masses above a few hundred GeV. For higher DM masses, the  limits on the DM annihilation cross section to muon pairs are obtained from the observation of the Milky Way Galactic halo by IceCube~\cite{Abbasi:2011eq}. In the same figure, we show the improvements that can be obtained by observing galaxy clusters using  neutrino telescopes. It is clear that galaxy clusters provide an order of magnitude more sensitivity compared to other sources. 

For this channel, we predict that the tracks observed in IceCube will give the best sensitivity for DM masses, $m_{\chi}\gsim 1$\,TeV. KM3NeT, augmented with a DeepCore-like low energy extension, will provide increased sensitivity to this annihilation cross section at DM masses, $m_{\chi}\lsim 100$\,GeV. Although the expected sensitivity is weaker than the limits obtained from Fermi-LAT observation of dwarf galaxies, it will be an important complementary test, as the neutrino observations are less dependent on the central density profile.

We show the various constraints on the $\chi\chi\,\rightarrow\,\nu\overline{\nu}$ channel on the right panel in Fig.\,\ref{fig:comparison}. This indirect detection channel can only be detected by neutrino telescopes and it has no signatures in any other DM indirect detection experiment.
  
For DM masses $m_{\chi}\gsim\,100$\,GeV, limits on the DM annihilation cross section to neutrino pairs are obtained from the observation of the Milky Way Galactic halo by IceCube~\cite{Abbasi:2011eq}. In the same figure, we show the improvements that can be obtained by observing galaxy clusters using  neutrino telescopes. Observation of the galaxy clusters by neutrino telescopes shall give an order of magnitude improvement over the existing constraints. 

For this channel, we predict that the observation of muon tracks in IceCube will give the best sensitivity above DM mass, $m_{\chi}\approx 300$\,GeV. KM3NeT augmented with a DeepCore-like low energy extension will provide the best sensitivity to this annihilation cross section at DM masses~$m_{\chi}\lsim 1$\,TeV. For DM masses~$\lsim 100$\,GeV, KM3NeT, with a low-energy extension, can reach annihilation cross sections of the order of $10^{-24}\,{\rm cm^3s^{-1}}$ while observing cascades in the detector. 

\section{Conclusion}       \label{sec:Conclusion}

In this paper, we have considered observation of galaxy clusters by neutrino telescopes and discussed the improvements that can be made over the existing limits. Recent high resolution computer simulations of galaxy clusters predict a large enhancement in the annihilation flux due to DM substructures. We take the substructure contribution into account and predict the neutrino flux from a typical galaxy cluster. We find that the sensitivity that can be obtained using galaxy clusters should improve the existing constraints by more than an order of magnitude. Our results should therefore encourage the IceCube collaboration to look at galaxy clusters, as an extension of their work on dwarf galaxies~\cite{IceCube:2011ae}.

Due to the extended nature of the DM substructure profile (see Fig.\,\ref{fig:J vs psi}), nearby galaxy clusters like Virgo should appear as extended sources at neutrino telescopes. We find that the optimal angular window around a galaxy cluster that maximizes the signal-to-noise ratio has a radius $\approx2^\circ$ (see Fig.\,\ref{fig:J over square root delta omega}).

An order of magnitude improvement over the IceCube sensitivity is expected if KM3NeT deploys a low energy extension (like DeepCore in IceCube) in their telescope, which would allow for a full-sky observation with good pointing using cascades. This has the potential to open the $(10\,{\rm GeV}-100\,{\rm GeV})$ DM mass range to neutrino astronomy, and improve existing constraints by an order of magnitude. We hope that these promising results will encourage the KM3NeT collaboration to investigate the possibility of deploying a low energy extension to their telescope and improve the reconstruction of cascades (see right panels in Fig.\,\ref{fig:sensitivity}).

We looked at the $\chi \chi \,\rightarrow \,\mu^+\mu^-$ annihilation channel and predicted an order of magnitude improvement over the current constraints (see left panel in Fig.\,\ref{fig:comparison}). Although this bound turns out to be weaker than the bound on the annihilation cross section given by Fermi-LAT while observing dwarf spheroidal galaxies, we emphasize that the large angular resolution of the neutrino telescopes make the result more model-independent than that obtained by Fermi-LAT. We have predicted that the improvement in sensitivity to the annihilation cross section in this channel will allow us to probe cross sections $\sigv\gsim(10^{-24}-10^{-22}){\rm cm^3s^{-1}}$ for DM masses in the range $(1\,{\rm GeV}-10\,{\rm TeV})$ for 10 years of observation by a km$^3$ neutrino telescope.

We have also looked at the $\chi \chi \,\rightarrow \, \nu\overline{\nu}$ channel and predicted that the observation of galaxy clusters will constrain the annihilation cross section in this channel by an order of magnitude over the existing limit obtained by IceCube while observing the Milky Way Galactic halo (see right panel in Fig.\,\ref{fig:comparison}). This annihilation channel is unique as it has no signal in any other DM indirect detection experiment. We predicted that the improvement in sensitivity to the annihilation cross section in this channel will allow us to probe cross sections $\sigv\gsim(10^{-24}-10^{-22}){\rm cm^3s^{-1}}$ for DM masses in the range $(1\,{\rm GeV}-10\,{\rm TeV})$ for 10 years of observation by a km$^3$ neutrino telescope.

We considered the $\chi \chi \,\rightarrow \, t\overline{t}$ annihilation channel, which is expected to be very important for a heavy fermionic DM particle. We have predicted that the improvement in sensitivity to the annihilation cross section in this channel will allow us to probe cross sections $\sigv\gsim10^{-22}{\rm cm^3s^{-1}}$ for 10 years of observation by a km$^3$ neutrino telescope.

We finally considered the $\chi \chi \,\rightarrow \, \nu\overline{\nu}\nu \overline{\nu}$ channel and predict that the sensitivity that can be obtained using neutrino telescopes may be able to probe the annihilation cross sections required in models which aim to solve various small-scale problems in $\Lambda$CDM.

Although we have performed our calculations for the Virgo galaxy cluster, we expect that neutrino telescope observation of a properly chosen galaxy cluster (after taking into consideration backgrounds and various detector systematics in more detail) will improve the limits on the annihilation cross section by an order of magnitude in almost all annihilation channels. We must emphasize that the biggest uncertainty in this result comes from the $\sim$11 orders of magnitude extrapolation in the minimum DM substructure mass that is used to calculate the DM substructure profile. As a consequence of this extrapolation of the minimum substructure mass, the boost factor that can be obtained in a galaxy cluster due to the presence of substructures can vary by a factor of $\sim$20. Unless simulations improve their resolution dramatically, this will remain an inherent assumption in any DM indirect detection experiment observing galaxy clusters.

All things considered, we hope to have conveyed the usefulness of observing galaxy clusters at neutrino telescopes for studying DM. In particular, how good reconstruction of cascades can lead to significant improvements in sensitivity. We hope that the IceCube and the KM3NeT collaborations will consider our results and make the required improvements in their analyses and detectors to make this possible. 

\section*{Acknowledgements}       \label{sec:acknowledgements}
We thank John Beacom and Kohta Murase for discussions about their related independent work~\cite{Murase:2012np}. We thank Shin'ichiro Ando, Shunsaku Horiuchi, Peter Melchior, Julian Merten, Kenny Chun Yu Ng,  Bjoern Opitz, and especially Carsten Rott for useful discussions and comments on the manuscript. We thank Joakim Edsjo, Savvas Koushiappas, Michael Kuhlen, and Emmanuel Moulin for discussions at the IDM2012 conference where this work was presented. We also thank the anonymous referee whose suggestions greatly improved the paper. R.\,L. is supported by NSF grant \mbox{PHY-1101216} awarded to John Beacom.

\end{document}